\documentclass[twocolumn]{aastex631}

\usepackage{amsmath}
\usepackage{amsfonts}
\usepackage{graphicx}
\usepackage{rotating}
\usepackage{natbib}
\usepackage{txfonts}
\usepackage{color}
\usepackage{wrapfig}
\usepackage{amssymb}
\usepackage{color}

\shorttitle{Hot core in the extreme outer Galaxy} 
\shortauthors{T. Shimonishi et al.} 

\begin{document}

\title{The detection of a hot molecular core in the extreme outer Galaxy}

\correspondingauthor{Takashi Shimonishi} 
\email{shimonishi@env.sc.niigata-u.ac.jp} 

\author{Takashi Shimonishi} 
\affiliation{Center for Transdisciplinary Research, Niigata University, Ikarashi-ninocho 8050, Nishi-ku, Niigata, 950-2181, Japan}
\affiliation{Environmental Science Program, Department of Science, Faculty of Science, Niigata University, Ikarashi-ninocho 8050, Nishi-ku, Niigata, 950-2181, Japan}

\author{Natsuko Izumi} 
\affiliation{Institute of Astronomy and Astrophysics, Academia Sinica, No. 1, Section 4, Roosevelt Road, Taipei 10617, Taiwan}

\author{Kenji Furuya} 
\affiliation{National Astronomical Observatory of Japan, Osawa 2-21-1, Mitaka, Tokyo 181-8588, Japan}

\author{Chikako Yasui} 
\affiliation{National Astronomical Observatory of Japan, California Office, 100 W. Walnut St., Suite 300, Pasadena, CA 91124, USA}



\begin{abstract}
Interstellar chemistry in low metallicity environments is crucial to understand chemical processes in the past metal-poor universe. 
\added{Recent studies of interstellar molecules in nearby low-metallicity galaxies have suggested that the metallicity has a significant effect on chemistry of star-forming cores. }
Here we report the first detection of a hot molecular core in the extreme outer Galaxy, which is an excellent laboratory to study star formation and interstellar medium in a Galactic low-metallicity environment. 
The target star-forming region, WB89-789, is located at the galactocentric distance of 19 kpc. \deleted{and the metallicity is a factor of four lower than the solar value. }
Our ALMA observations \added{in 241-246, 256-261, 337-341, and 349-353 GHz} have detected a variety of carbon-, oxygen-, nitrogen-, sulfur-, and silicon-bearing species, including complex organic molecules (COMs) containing up to nine atoms, towards a warm ($>$100 K) and compact ($<$ 0.03 pc) region associated with a protostar ($\sim$8 $\times$ 10$^3$ L$_{\sun}$). 
\added{Deuterated species such as HDO, HDCO, D$_2$CO, and CH$_2$DOH are also detected. }
A comparison of fractional abundances of COMs relative to CH$_3$OH between the outer Galactic hot core and an inner Galactic counterpart shows a remarkable similarity. 
On the other hand, the molecular abundances in the present source do not resemble those of low-metallicity hot cores in the \replaced{LMC}{Large Magellanic Cloud}. 
\added{The results suggest that a great molecular complexity exists even in a primordial environment of the extreme outer Galaxy. }
The detection of another embedded protostar associated with high-velocity SiO outflows is also reported. 
\end{abstract} 

\keywords{astrochemistry -- ISM: molecules -- stars: protostars -- ISM: jets and outflows -- radio lines: ISM}



\section{Introduction} \label{sec_intro} 
Understanding the star formation and interstellar medium (ISM) at low metallicity is crucial to unveil physical and chemical processes in the past Galactic environment or those in high-redshift galaxies, where the metallicity was significantly lower compared to the present-day solar neighborhood. 

Hot \added{molecular} cores are one of the early stages of star formation and they play a key role in the formation of chemical complexity of the ISM. 
Physically, hot cores are defined as having small source size (\replaced{$\leq$}{$\lesssim$}0.1 pc), high density (\replaced{$\geq$}{$\gtrsim$}10$^6$ cm$^{-3}$), and warm gas/dust temperature (\replaced{$\geq$}{$\gtrsim$}100 K) \citep[e.g., ][]{vanD98, Kur00}. 
Chemistry of hot cores is characterized by sublimation of ice mantles, which accumulated in the course of star formation. 
In cold molecular clouds and prestellar cores, gaseous molecules and atoms are frozen onto dust grains. 
With increasing dust temperatures by star formation activities, chemical reaction among heavy species become active on grain surfaces to form larger complex molecules \citep[e.g., ][]{Gar06}. 
In addition, sublimated molecules, such as CH$_3$OH and NH$_3$, are subject to further gas-phase reactions \citep[e.g., ][]{NM04,Taq16}. 
As a result, warm and dense gas around protostars become chemically rich, and embedded protostars are observed as one of the most powerful molecular line emitters, which is called a hot core. 
They are important targets for astrochemical studies of star-forming regions, because a variety of molecular species, including complex organic molecules (COMs), are often detected in hot cores \citep[][and references therein]{Her09}. 
Thus detailed studies on chemical properties of hot cores are important for understanding complex chemical processes triggered by star formation. 

Recent ALMA (Atacama Large Millimeter/submillimeter Array) observations of hot molecular cores in a nearby low metallicity galaxy, the Large Magellanic Cloud (LMC), have suggested that the metallicity has a significant effect on their molecular compositions \citep{ST16b, ST20, Sew18}; cf., the metallicity of the LMC is $\sim$1/2-1/3 of the solar neighborhood. 
A comparison of molecular abundances between LMC and Galactic hot cores suggests that organic molecules (e.g., CH$_3$OH, a classical hot core tracer) show a large abundance variation in low-metallicity hot cores \citep{ST20}. 
There are organic-poor hot cores that are unique in the LMC \citep{ST16b}, while there are relatively organic-rich hot cores, where the abundances of organic molecules roughly scale with the metallicity \citep{Sew18}. 
Astrochemical simulations for low-metallicity hot cores suggest that dust temperature during the initial ice-forming stage would play a key role for making the chemical diversity of organic molecules \citep{Ach18, ST20}. 
On the other hand, sulfur-bearing molecules such as SO$_2$ and SO are commonly detected in known LMC hot cores and their molecular abundances \replaced{simply}{roughly} scale with the metallicity of the LMC. 
Although the reason is still under debate, the results suggest that SO$_2$ can be an alternative molecular species to trace hot core chemistry in metal-poor environments. 

The above results suggest that molecular abundances in hot cores do not always simply scale with the elemental abundances of their parent environments. 
However, it is still unclear if the observed chemical characteristics of LMC hot cores are common in other low metallicity environments or they are uniquely seen only in the LMC. 
Currently, known low-metallicity hot core samples are limited to those in the LMC. 
It is thus vital to understand universal characteristics of interstellar chemistry by studying chemical compositions of star-forming cores in diverse metallicity environments. 

Recent surveys \citep[e.g.,][]{And15,And18,Izu17,Wen21} have found a number of ($\sim$10-20) star-forming region candidates in the extreme outer Galaxy, which is defined as having galactocentric distance ($D_{GC}$) larger than 18 kpc \citep{Yas06,Kob08}. 
The extreme outer Galaxy has a very different environment from those in the solar neighborhood, with lower metallicity \citep[less than -0.5 dex,][]{Fer17,Wen19}, lower gas density \citep[e.g.,][]{Nak16}, and small or no perturbation from spiral arms. 
Such an environment is of great interest for studies of the star formation and ISM in the early phase of the Milky Way formation and those in dwarf galaxies \citep{Fer98,Kob08}. 
The low metallicity environment is in common with the Magellanic Clouds, and thus the extreme outer Galaxy is an ideal laboratory to test the universality of the low metallicity molecular chemistry observed in the LMC and SMC. 

Among star-forming regions in the extreme outer Galaxy, WB89-789 (IRAS 06145+1455; 06$^\mathrm{h}$17$^\mathrm{m}$24$\fs$2, 14$\arcdeg$54$\arcmin$42$\arcsec$, J2000) has particularly young and active nature \citep{Bra94}. 
It is located at the galactocentric distance of 19.0 kpc and the distance from Earth is 10.7 kpc \citep[based on optical spectroscopy of a K3 III star,][]{Bra07}. 
The metallicity of WB89-789 is estimated to be a factor of four lower than the solar value according to the Galactic oxygen abundance gradient reported in the literature \citep{Fer17,Wen19,Bra19,Are20,Are21}. 
The region is associated with dense clouds traced by CS and CO \citep{Bra07}. 
The total mass of the cloud is estimated to be 6 $\times$ 10$^3$ M$_{\sun}$ for a $\sim$10 pc diameter area \citep{Bra94}. 
An H$_2$O maser is detected towards the region \citep{Wou93}, but no centimeter radio continuum is found \citep{Bra07}. 
Several class I protostar candidates are identified by previous infrared observations \citep{Bra07}. 

We here report the first detection of a hot molecular core in the extreme outer Galaxy based on submillimeter observations towards WB89-789 with ALMA. 
Section \ref{sec_tarobsred} describes the details of the target source, observations, and data reduction. 
The observed molecular line spectra and images, as well as analyses of physical and chemical properties of the source, are presented in Section \ref{sec_res}. 
Discussion about the properties of the hot core and comparisons of molecular abundances with known Galactic and LMC hot cores are given in Section \ref{sec_disc}. 
This section also presents the detection of another embedded protostar with high-velocity outflows in the WB89-789 region. 
The conclusions are given in Section \ref{sec_sum}.

\begin{deluxetable*}{ l c c c c c c c c c}
\tablecaption{Observation summary \label{tab_Obs}} 
\tablewidth{0pt} 
\tabletypesize{\footnotesize} 
\tablehead{
\colhead{}   & \colhead{Observation} &  \colhead{On-source} & \colhead{Mean}                             & \colhead{Number}   &  \multicolumn{2}{c}{Baseline}     &  \colhead{}                                                  &  \colhead{}                                         &  \colhead{Channel}  \\
\cline{5-6}  
\colhead{}   & \colhead{Date}             &  \colhead{Time}         &  \colhead{PWV\tablenotemark{a}} & \colhead{of}              &  \colhead{Min} & \colhead{Max} & \colhead{Bem size\tablenotemark{b}}        & \colhead{MRS\tablenotemark{c}}     &  \colhead{Spacing}  \\
\colhead{}   & \colhead{}                    &  \colhead{(min)}         &  \colhead{(mm)}                             & \colhead{Antennas} & \colhead{(m)} & \colhead{(m)}   &  \colhead{($\arcsec$ $\times$ $\arcsec$)} &  \colhead{($\arcsec$)}                       &  \colhead{}                }
\startdata 
Band 6       &  2018 Dec 6 --             &  115.5                          &  0.5--1.5                                         &  45--49                     &  15.1               &  783.5              &  0.41 $\times$ 0.50                                     &   5.6                              &  0.98 MHz               \\
(250 GHz)  &  2019 Apr 16               &                                    &                                                       &                                  &                        &                         &                                                                     &                                      &  (1.2 km s$^{-1}$)    \\
Band 7       &  2018 Apr 30 --            &  64.1                           &  0.6--1.0                                         &  43-44                       &  15.1               &  500.2              &  0.46 $\times$ 0.52                                     &   5.4                              &  0.98 MHz                \\
(350 GHz)  &  2018 Aug 22              &                                    &                                                       &                                  &                        &                         &                                                                     &                                      &  (0.85 km s$^{-1}$)    \\
\enddata
\tablenotetext{a}{Precipitable water vapor.}
\tablenotetext{b}{The average beam size of continuum achieved by TCLEAN with the Briggs weighting and the robustness parameter of 0.5. 
Note that we use a common circular restoring beam size of 0$\farcs$50 for Band 6 and 7 data to construct the final images.}
\tablenotetext{c}{Maximum Recoverable Scale.}
\end{deluxetable*}

\section{Target, observations, and data reduction} \label{sec_tarobsred} 
\subsection{Target} \label{sec_tar}
\begin{figure}[tp] 
\begin{center} 
\includegraphics[width=5.5cm]{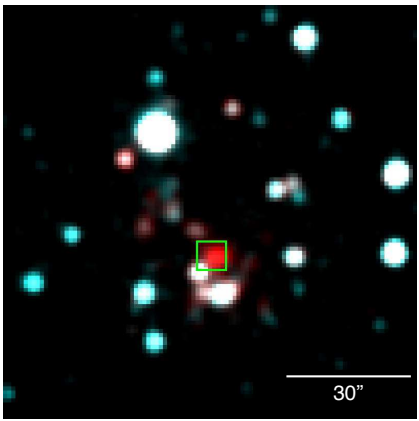} 
\caption{
Near-infrared two-color image of the WB89-789 star-forming region based on 2MASS data \citep{Skr06}. 
Blue is $J$-band (1.25 $\mu$m) and red is $K_s$-band (2.16 $\mu$m). 
The image size is 100\arcsec $\times$ 100\arcsec. 
The green square indicates the field-of-view of the ALMA submillimeter images shown in Figures \ref{images1}--\ref{images2}. 
}
\label{IR_image}
\end{center}
\end{figure}

The target star-forming region is WB89-789 \citep{Bra94}. 
The region contains three Class I protostar candidates identified by near-infrared observations \citep{Bra07}, and one of them is a main target of the present ALMA observations. 
The region observed with ALMA is indicated on a near-infrared two-color image shown in Figure \ref{IR_image}. 
The observed position is notably reddened compared with other parts of WB89-789. 

\subsection{Observations} \label{sec_obs} 
Observations were conducted with ALMA in 2018 and 2019 as a part of the Cycle 5 (2017.1.01002.S) and Cycle 6 (2018.1.00627.S) programs (PI: T. Shimonishi). 
A summary of the present observations is shown in Table \ref{tab_Obs}. 
The pointing center of antennas is RA = 06$^\mathrm{h}$17$^\mathrm{m}$23$^\mathrm{s}$ and Dec = 14$\arcdeg$54$\arcmin$41$\arcsec$ (ICRS). 
The total on-source integration time is 115.5 minutes for Band 6 data and 64.1 minutes for Band 7. 
Flux and bandpass calibrators are J0510+1800, J0854+2006, and J0725-0054 for Band 6, while J0854+2006 and J0510+1800 for Band 7, respectively. 
Phase calibrators are J0631+2020 and J0613+1708 for Band 6 and J0643+0857 and J0359+1433 for Band 7. 
Four spectral windows are used to cover the sky frequencies of 241.40--243.31, 243.76-245.66, 256.90--258.81, and 258.76--260.66 GHz for Band 6, while 337.22--339.15, 339.03-340.96, 349.12--351.05, and 350.92--352.85 GHz for Band 7. 
The channel spacing is 0.98 MHz, which corresponds to 1.2 km s$^{-1}$ for Band 6 and 0.85 km s$^{-1}$ for Band 7. 
The total number of antennas is 45--49 for Band 6 and 43--44 for Band 7. 
The minimum--maximum baseline lengths are 15.1--783.5 m for Band 6 and 15.1--500.2 m for Band 7. 
A full-width at half-maximum (FWHM) of the primary beam is about 25$\arcsec$ for Band 6 and 18$\arcsec$ for Band 7.

\subsection{Data reduction} \label{sec_red} 
Raw data is processed with the \textit{Common Astronomy Software Applications} (CASA) package. 
We use CASA 5.4.0 (Band 6) and 5.1.1 (Band 7) for the calibration and CASA 5.5.0 for the imaging. 
The synthesized beam sizes of 0$\farcs$39--0$\farcs$42 $\times$ 0$\farcs$49--0$\farcs$52 with a position angle of -36 degree for Band 6 and 0$\farcs$45--0$\farcs$46 $\times$ 0$\farcs$51--0$\farcs$52 with a position angle of - 54 degree for Band 7 are achieved with the Briggs weighting and the robustness parameter of 0.5. 
In this paper, we use a common circular restoring beam size of 0$\farcs$50, which corresponds to 0.026 pc (5350 au) at the distance of WB89-789. 
The synthesized images are corrected for the primary beam pattern using the impbcor task in CASA. 
The continuum image is constructed by selecting line-free channels. 
Before the spectral extraction, the continuum emission is subtracted from the spectral data using the CASA's uvcontsub task. 

The spectra and continuum flux are extracted from the 0$\farcs$50 diameter circular region centered at RA = 06$^\mathrm{h}$17$^\mathrm{m}$24$\fs$073 and Dec = 14$\arcdeg$54$\arcmin$42$\farcs$27 (ICRS), which corresponds to the submillimeter continuum center of the target and is equivalent to the hot core position. 
Hereafter, the source is referred to as WB89-789 SMM1. 
\deleted{The extracted spectra are shown in Figures \ref{spec_B6}--\ref{spec_B7}. }

\begin{deluxetable*}{ l l l l l l l l }[tbp!]
\tablecaption{Summary of detected molecular species \label{tab_line_summary}}
\tablewidth{0pt}
\tabletypesize{\small} 
\tablehead{
\colhead{2 atoms}   & \colhead{3 atoms}   &       \colhead{4 atoms}   &    \colhead{5 atoms}&       \colhead{6 atoms}  &       \colhead{7 atoms} &       \colhead{8 atoms} &       \colhead{9 atoms}            \\
}
\startdata 
CN                          &   HDO                       &       H$_2$CO               &       c-C$_3$H$_2$       &           CH$_3$OH       &         CH$_3$CHO    &  HCOOCH$_3$          &   CH$_3$OCH$_3$          \\
NO                          &   H$^{13}$CO$^+$   &      HDCO                     &        HC$_3$N              &     $^{13}$CH$_3$OH  &     c-C$_2$H$_4$O   &                                   &    C$_2$H$_5$OH                \\
CS                          &    HC$^{18}$O$^+$  &      D$_2$CO                &         H$_2$CCO          &         CH$_2$DOH      &                                    &                                   &   C$_2$H$_5$CN          \\
C$^{34}$S              &   H$^{13}$CN           &       HNCO                    &         HCOOH               &          CH$_3$CN         &                                   &                                    &       \\
C$^{33}$S              &   HC$^{15}$N           &       H$_2$CS               &                                      &            NH$_2$CO        &                                    &                                    &       \\
SO                          &   CCH                       &                                        &                                     &                                    &                                      &                                    &       \\
$^{34}$SO              &  SO$_2$                   &                                      &                                    &                                    &                                      &                                    &       \\
$^{33}$SO              &   $^{34}$SO$_2$      &                                     &                                    &                                    &                                      &                                    &       \\
SiO                         &   OCS                       &                                     &                                     &                                    &                                      &                                    &       \\
                               &   $^{13}$OCS           &                                     &                                      &                                    &                                      &                                    &       \\
\enddata
\end{deluxetable*}

\section{Results and analysis} \label{sec_res} 
\subsection{Spectra} \label{sec_spc} 
\begin{figure*}[tp!] 
\begin{center} 
\includegraphics[width=17cm]{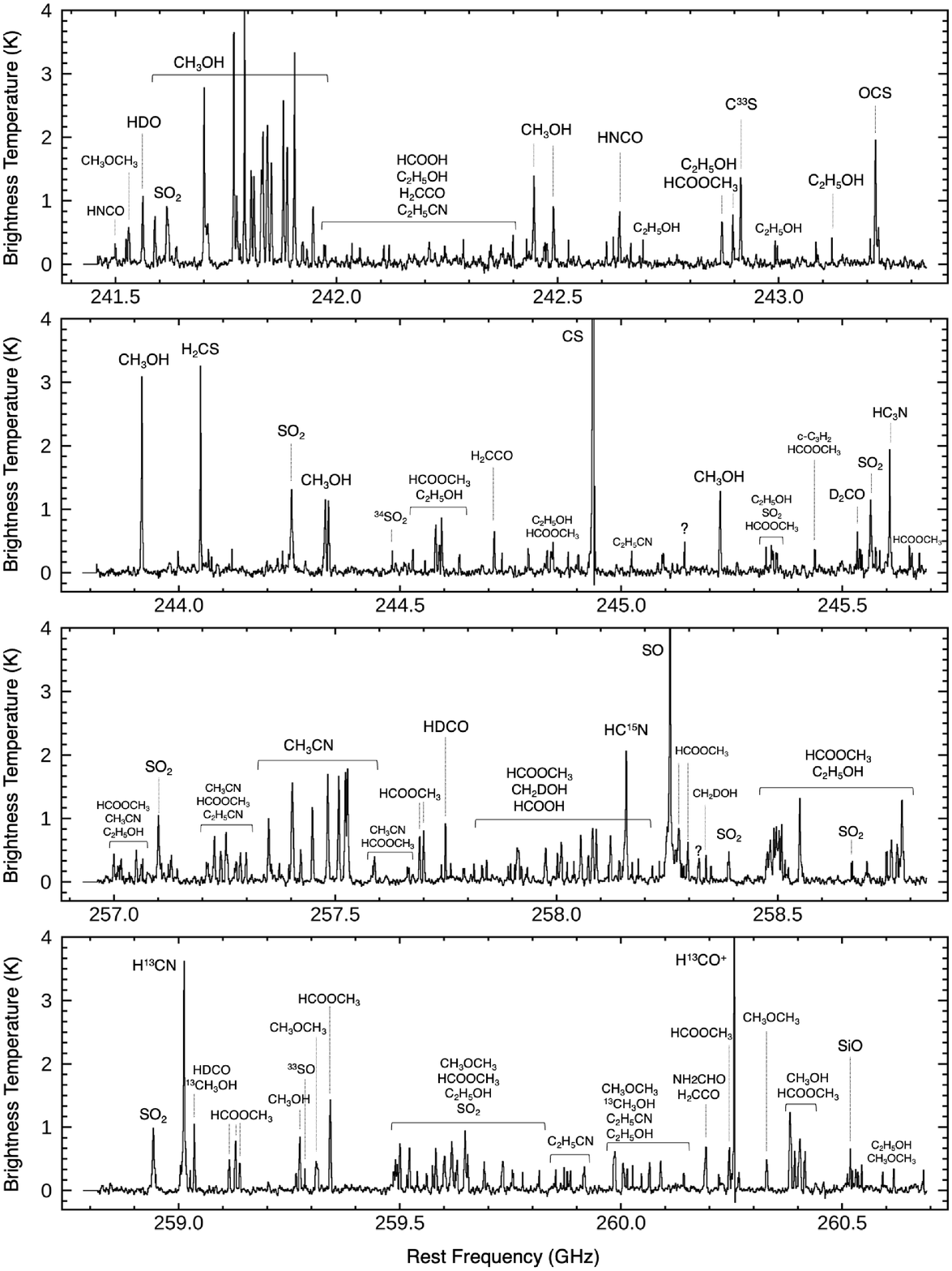} 
\caption{
ALMA band 6 spectra extracted from the the 0$\farcs$50 (0.026 pc) diameter region centered at the present hot molecular core in the extreme outer Galaxy, WB89-789 SMM1. 
Detected emission lines are labeled. 
Unidentified lines are indicated by ``?". 
The source velocity of 34.5 km s$^{-1}$ is assumed. 
}
\label{spec_B6}
\end{center}
\end{figure*}

\begin{figure*}[tp!] 
\begin{center} 
\includegraphics[width=17cm]{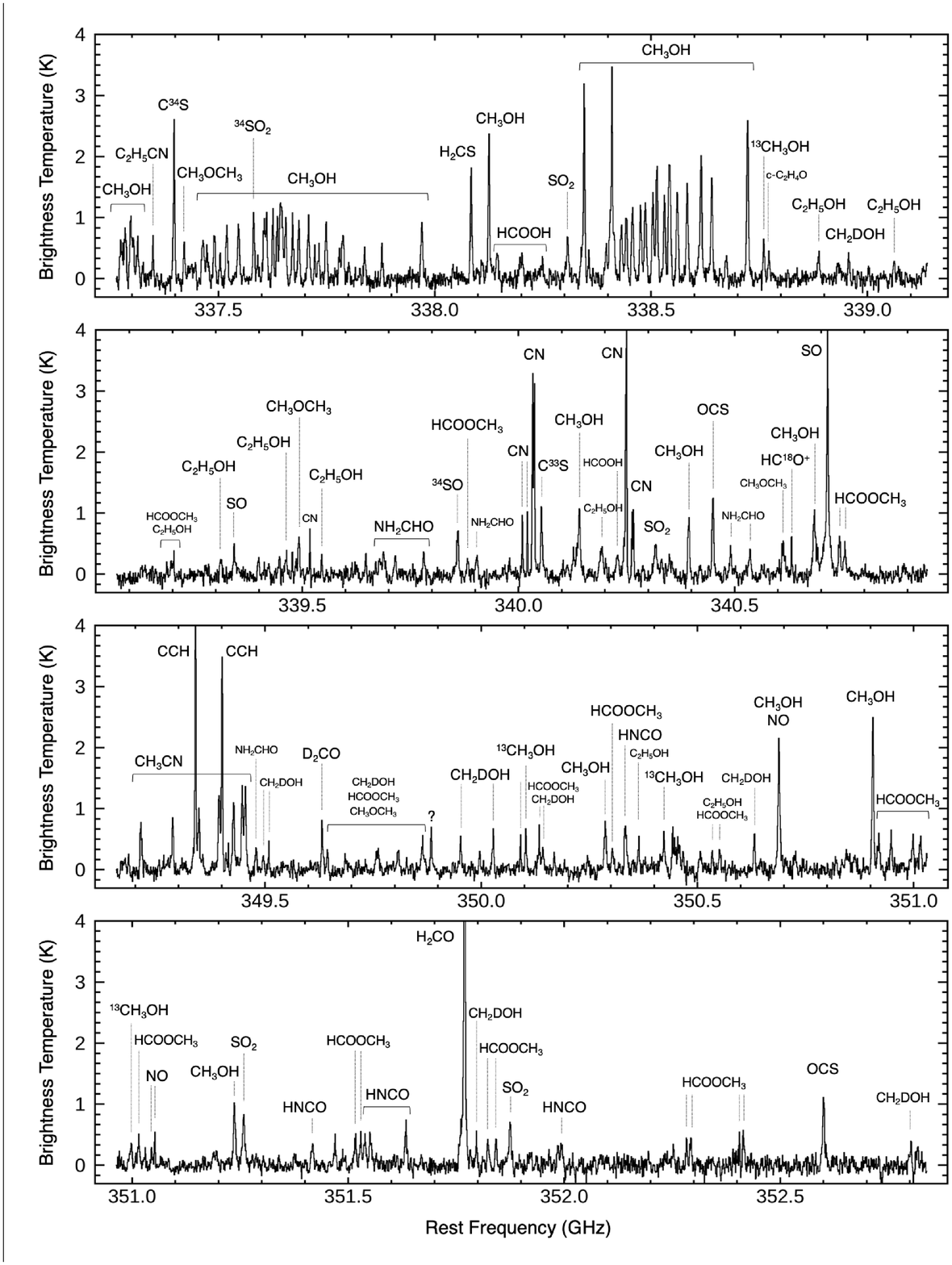} 
\caption{
Same as in Figure \ref{spec_B6}, but for ALMA Band 7. 
}
\label{spec_B7}
\end{center}
\end{figure*}

Figures \ref{spec_B6}--\ref{spec_B7} show submillimeter spectra extracted from the continuum center of WB89-789 SMM1. 
Spectral lines are identified with the aid of the Cologne Database for Molecular Spectroscopy\footnote{https://www.astro.uni-koeln.de/cdms} \citep[CDMS,][]{Mul01,Mul05} and the molecular database of the Jet Propulsion Laboratory\footnote{http://spec.jpl.nasa.gov} \citep[JPL,][]{Pic98}. 
\added{The detection criterion adopted here is 3$\sigma$ significance level and the velocity coincidence with the systemic velocity ($V_{sys}$) of WB89-789 SMM1 (34.5 km s$^{-1}$). 
The lines with the significance level higher than 2.5$\sigma$ but lower than 3$\sigma$ are indicated as tentative detection in the tables in Appendix A. 
More than 85 $\%$ of lines are detected above 5$\sigma$ level. } 

Line parameters are measured by fitting a Gaussian profile to detected lines. 
We estimate the peak brightness temperature, the FWHM, the LSR velocity, and the integrated intensity for each line based on the fitting. 
For spectral lines for which a Gaussian profile does not fit well, their integrated intensities are calculated by directly integrating the spectrum over the frequency region of emission. 
Full details of the line fitting can be found in Appendix A (Tables of measured line parameters) and Appendix B (Figures of fitted spectra). 
The table also contains the estimated upper limits on important non-detection lines. 

A variety of carbon-, oxygen-, nitrogen-, sulfur-, and silicon-bearing species, including COMs containing up to nine atoms, are detected from WB89-789 SMM1 (see Table \ref{tab_line_summary}). 
Multiple high excitation lines (upper state energy $>$100 K) are detected for many species. 
Measured line widths are typically 3--6 km s$^{-1}$. 
Most of lines consist of a single velocity component, but SiO has doppler shifted components at $V_{sys}$ $\pm$ 5 km s$^{-1}$ as indicated in Figure \ref{line_others} in Appendix B. 

\subsection{Images} \label{sec_img} 
\begin{figure*}[tp!]
\begin{center}
\includegraphics[width=17.5cm]{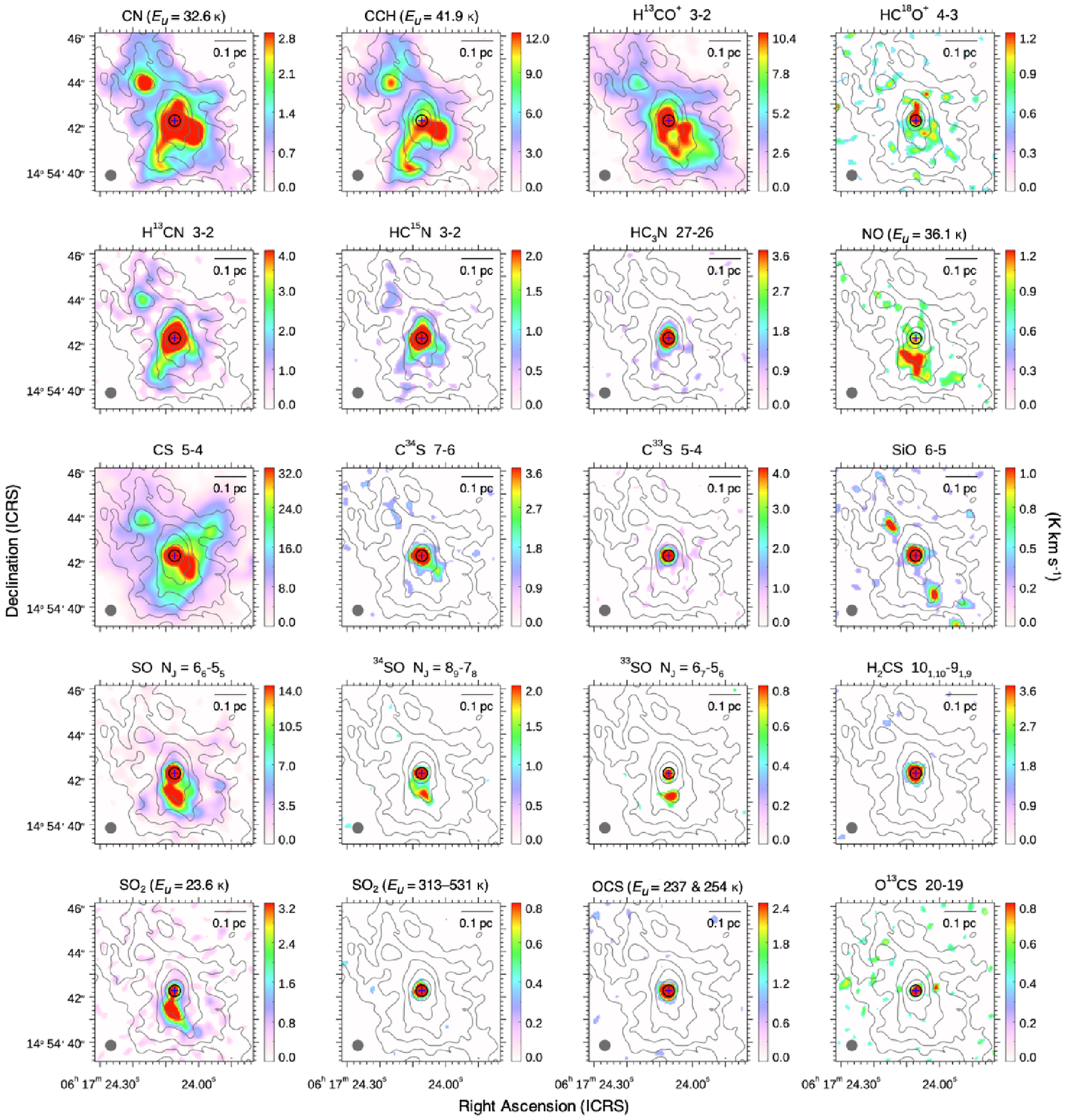}
\caption{
Integrated intensity distributions of molecular emission lines. 
Gray contours represent the 1.2 mm continuum distribution and the contour levels are 5$\sigma$, 10$\sigma$, 20$\sigma$, 40$\sigma$, 100$\sigma$ of the rms noise (0.044 mJy/beam). 
Low signal-to-noise ratio regions (S/N $<$2) are masked. 
The spectra discussed in the text are extracted from the region indicated by the black open circle. 
The blue cross represents the 1.2 mm continuum center. 
The synthesized beam size is shown by the gray filled circle in each panel. 
North is up, and east is to the left. 
}
\label{images1}
\end{center}
\end{figure*}

\begin{figure*}[tp!]
\begin{center}
\includegraphics[width=17.5cm]{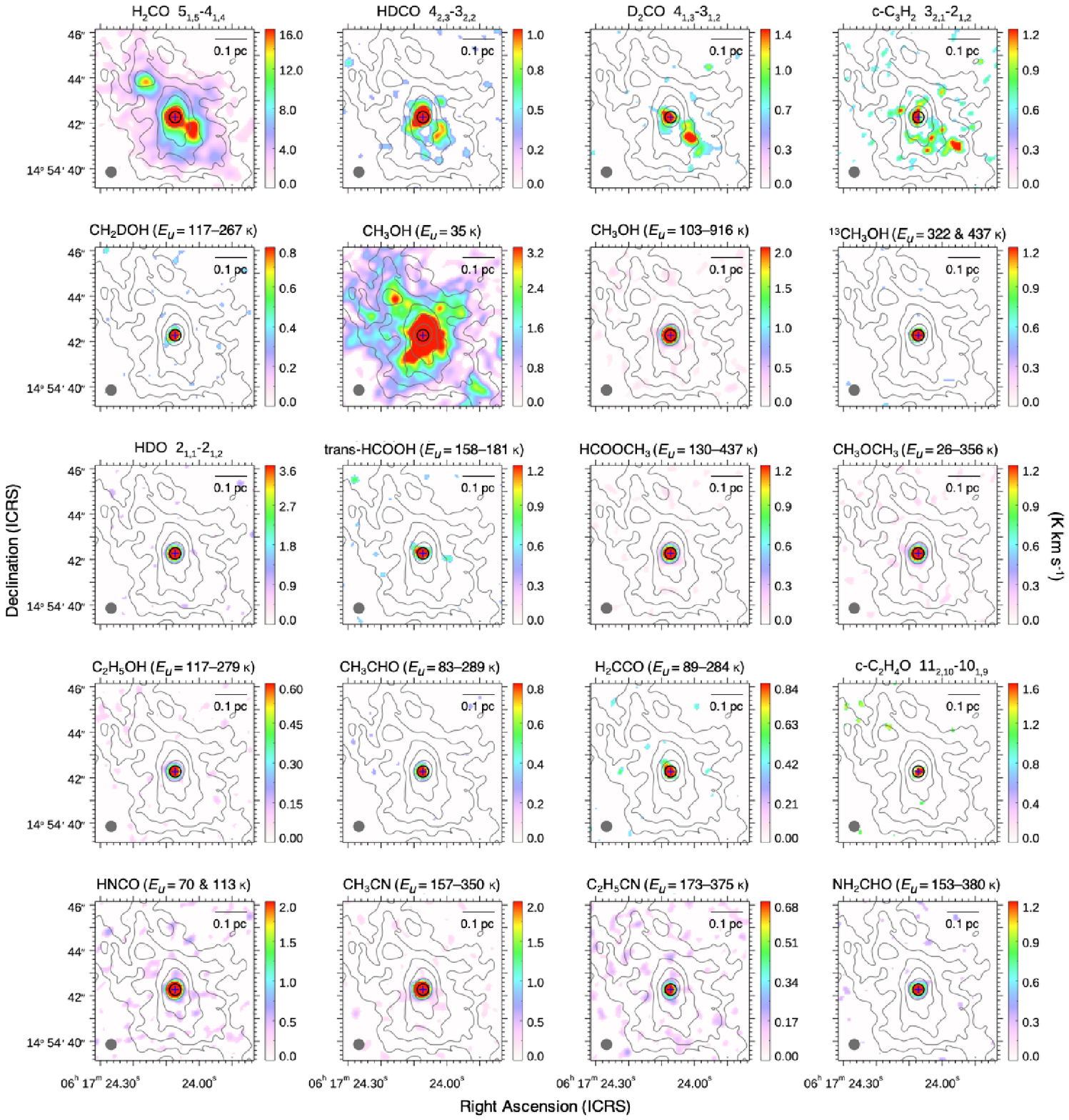}
\caption{
Same as in Figure \ref{images1}. 
}
\label{images2}
\end{center}
\end{figure*}

Figures \ref{images1}--\ref{images2} show synthesized images of continuum and molecular emission lines observed toward the target region. 
The images are constructed by integrating spectral data in the velocity range where the emission is detected. 
Most molecular lines, except for those of molecular radicals CN, CCH, and NO, have their intensity peak at the continuum center, which corresponds to the position of a hot core. 
Simple molecules such as H$^{13}$CO$^+$, H$^{13}$CN, CS, and SO are extended compared to the beam size. 
Secondary intensity peaks are also seen in those species. 
Complex molecules and HDO are concentrated at the hot core position. 
A characteristic symmetric distribution is seen in SiO. 
Further discussion about the distribution of the observed emission is presented in Section \ref{sec_disc_dist}. 

\subsection{Derivation of column densities, gas temperatures, and molecular abundances} \label{sec_ana}
\subsubsection{Rotation diagram analysis} \label{sec_rd}
\begin{figure*}[tp!]
\begin{center}
\includegraphics[width=16cm]{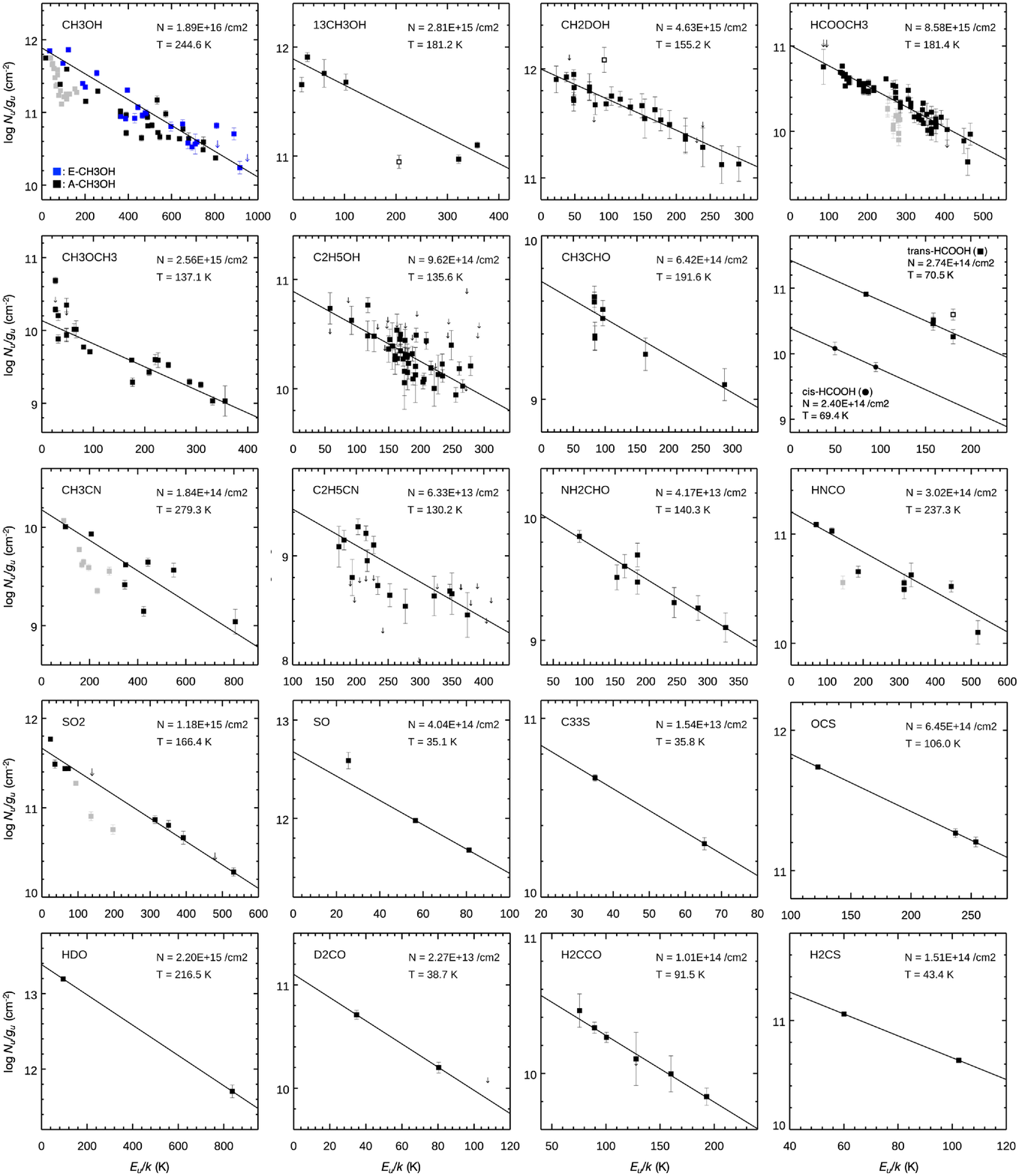}
\caption{
Results of rotation diagram analyses. 
Upper limit points are shown by the downward arrows. 
The solid lines represent the fitted straight line. 
Derived column densities and rotation temperatures are shown in each panel. 
The open squares are excluded in the fit because they significantly deviate from other data points. 
The gray squares are also excluded in the fit because of their large $S\mu^2$ values. 
CH$_3$OH is fitted by using only E-type transitions, which are shown in blue. 
For HCOOH, trans- (square) and cis- (circle) species are plotted together. 
See Section \ref{sec_rd} for details. 
}
\label{rd1}
\end{center}
\end{figure*}

Column densities and rotation temperatures are estimated based on the rotation diagram analysis for the molecular species where multiple transitions with different excitation energies are detected (Figure \ref{rd1}). 
We here assume an optically thin condition and the local thermodynamic equilibrium (LTE). 
We use the following formulae based on the standard treatment of the rotation diagram analysis \citep[e.g., ][]{Sut95, Gol99}: 
\begin{equation}
\log \left(\frac{ N_{u} }{ g_{u} } \right) = - \left(\frac {\log e}{T_{\mathrm{rot}}} \right) \left(\frac{E_{u}}{k} \right) + \log \left(\frac{N}{Q(T_{\mathrm{rot}})} \right),  \label{Eq_rd1}
\end{equation}
where 
\begin{equation}
\frac{ N_{u} }{ g_{u} } = \frac{ 3 k \int T_{\mathrm{b}} dV }{ 8 \pi^{3} \nu S \mu^{2} }, \label{Eq_rd2} \\ 
\end{equation}
and $N_{u}$ is a column density of molecules in the upper energy level, $g_{u}$ is the degeneracy of the upper level, $k$ is the Boltzmann constant, $\int T_{\mathrm{b}} dV$ is the integrated intensity estimated from the observations, $\nu$ is the transition frequency, $S$ is the line strength, $\mu$ is the dipole moment, $T_{\mathrm{rot}}$ is the rotational temperature, $E_{u}$ is the upper state energy, $N$ is the total column density, and $Q(T_{\mathrm{rot}})$ is the partition function at $T_{\mathrm{rot}}$. 
All the spectroscopic parameters required in the analysis are extracted from the CDMS or JPL database. 
Derived column densities and rotation temperatures are summarized in Table \ref{tab_N}. 

Most molecular species are well fitted by a single temperature component. 
Data points in diagrams of CH$_3$CN and C$_2$H$_5$CN are relatively scattered. 
For CH$_3$OH, CH$_3$CN, HNCO, SO$_2$, and HCOOCH$_3$, transitions with relatively large $S\mu^2$ values at low $E_{u}$ ($<$300 K) are excluded from the fit in order to avoid possible effect of optical thickness (see gray points in Fig. \replaced{\ref{sec_rd}}{\ref{rd1}}). 
\added{Adapted threshold values are log $S\mu^2$ $>$1.1 for CH$_3$OH, log $S\mu^2$ $>$2.4 for CH$_3$CN, log $S\mu^2$ $>$1.6 for HNCO, log $S\mu^2$ $>$1.2 for SO$_2$, and log $S\mu^2$ $>$1.8 for HCOOCH$_3$. }

Complex organic molecules, HDO, and SO$_2$ show high rotation temperatures ($>$130 K). 
This suggests that they are originated from a warm region associated with a protostar. 
On the other hand, C$^{33}$S and D$_2$CO, and H$_2$CS show lower temperatures, suggesting that they arise from a colder region in the outer part of the protostellar envelope. 
\added{SO also shows a low rotation temperature. 
Its $T_{\mathrm{rot}}$ is close to that of C$^{33}$S. 
However, SO lines are often optically thick in dense cores, particularly for low-$E_{u}$ lines, thus the derived rotation temperature would be an upper limit. }

\subsubsection{Column densities of other molecules} \label{sec_n} 
Column densities of molecular species for which rotation diagram analysis is not applicable are estimated from Equation \ref{Eq_rd1} after solving it for $N$. 
Their rotation temperatures are estimated as follows, by taking into account that the sight-line of WB89-789 SMM1 contains both cold and warm gas components as described in Section \ref{sec_rd}. 

The rotation temperature of C$^{33}$S is applied to those of CS and C$^{34}$S, considering a similar distribution of isotopologues. 
Similarly, the rotation temperature of D$_2$CO is applied to H$_2$CO and HDCO, and that of SO$_2$ to $^{34}$SO$_2$. 
For other species, we assume that molecules with an extended spatial distribution trace a relatively low-temperature region rather than a high-temperature gas associated with a hot core. 
CN, CCH, H$^{13}$CO$^+$, HC$^{18}$O$^+$, H$^{13}$CN, HC$^{15}$N, NO, SiO\deleted{, SO}, $^{34}$SO, $^{33}$SO, and c-C$_3$H$_2$ correspond to this case. 
We assume a rotation temperature of 35 K for those species, which is roughly equivalent to that of C$^{33}$S. 

High gas temperatures are observed for COMs, SO$_2$, and HDO, which are associated with a compact hot core region. 
Average temperature of those species is $\sim$200 K. 
We assume this temperature for column density estimates (including upper limit) of c-C$_2$H$_4$O, HC$_3$N, $^{13}$CH$_3$CN, $^{13}$OCS, and CH$_3$SH. 
Estimated column densities are summarized in Table \ref{tab_N}. 

We have also estimated column densities of selected species based on non-LTE calculations with RADEX \citep{vdT07}. 
For input parameters, we use the H$_2$ gas density of 2.1 $\times$ 10$^7$ cm$^{-3}$ according to our estimate in Section \ref{sec_h2} and the background temperature of 2.73 K. 
Kinetic temperatures are assumed to be the same as temperatures tabulated in Table \ref{tab_N}. 
The line intensities and widths are taken from the tables in Appendix A \footnote{The following lines are used for non-LTE calculation with RADEX; H$^{13}$CO$^+$(3--2), HC$^{18}$O$^+$(4--3), H$_2$CO(5$_{1,5}$--4$_{1,4}$), c-C$_3$H$_2$(3$_{2,1}$--2$_{1,2}$), CN(N = 3--2, J = $\frac{5}{2}$--$\frac{3}{2}$, F = $\frac{5}{2}$--$\frac{5}{2}$),  H$^{13}$CN(3--2), HC$^{15}$N(3--2), HC$_3$N(27--26), NO(J = $\frac{7}{2}$--$\frac{5}{2}$, $\Omega$ = $\frac{1}{2}$, F = $\frac{9}{2}$$^+$--$\frac{7}{2}$$^-$), CH$_3$CN(14$_{0}$--13$_{0}$), SiO(6--5), CS(5--4), OCS(20--19), H$_2$CS(7$_{1,6}$--6$_{1,5}$), SO($N_J$ = 6$_{6}$--5$_{5}$), and CH$_3$OH(7$_{5}$ E--6$_{5}$ E). }. 
We assume an empirical 10$\%$ uncertainty for input line intensities. 
The resultant column densities are summarized in Table \ref{tab_N}. 
The calculated non-LTE column densities are reasonably consistent with the LTE estimates.

\subsubsection{Column density of H$_2$, dust extinction, and gas mass} \label{sec_h2} 
A column density of molecular hydrogen ($N_{\mathrm{H_2}}$) is estimated from the dust continuum data. 
We use the following equation to calculate $N_{\mathrm{H_2}}$ based on the standard treatment of optically thin dust emission: 
\begin{equation}
N_{\mathrm{H_2}} = \frac{F_{\nu} / \Omega}{2 \kappa_{\nu} B_{\nu}(T_{d}) Z \mu m_{\mathrm{H}}} \label{Eq_h2}, 
\end{equation}
where $F_{\nu}/\Omega$ is the continuum flux density per beam solid angle as estimated from the observations, $\kappa_{\nu}$ is the mass absorption coefficient of dust grains coated by thin ice mantles at 1200/870 $\mu$m as taken from \citet{Oss94} and we here use 1.07 cm$^2$ g$^{-1}$ for 1200 $\mu$m and 1.90 cm$^2$ g$^{-1}$ for 870 $\mu$m, $T_{d}$ is the dust temperature and $B_{\nu}(T_{d})$ is the Planck function, $Z$ is the dust-to-gas mass ratio, $\mu$ is the mean atomic mass per hydrogen \citep[1.41, according to][]{Cox00}, and $m_{\mathrm{H}}$ is the hydrogen mass. 
We use the dust-to-gas mass ratio of 0.002, which is obtained by scaling the Galactic value of 0.008 by the metallicity of the WB89-789 region. 

A line of sight towards a hot core contain dust grains with different temperatures because of the temperature gradient in a protostellar envelope. 
Representative dust temperature (i.e. mass-weighted average temperature) would fall somewhere in between that of a warm inner region and a cold outer region. 
\citet{ST20} presented a detailed analysis of effective dust temperature in the sight-line of a low-metallicity hot core in the LMC, based on a comparison of $N_{\mathrm{H_2}}$ derived by submillimeter dust continuum with the above method, model fitting of spectral energy distributions (SEDs), and the 9.7 $\mu$m silicate dust absorption depth. 
The paper concluded that $T_{d}$ = 60 K for the dust continuum analysis yields the $N_{\mathrm{H_2}}$ value which is consistent with those obtained by other different methods. 
This temperature corresponds to an intermediate value between a cold gas component ($\sim$50 K) represented by SO and a warm component ($\sim$150 K) represented by CH$_3$OH and SO$_2$ in this LMC hot core. 
The present hot core, WB89-789 SMM1, harbors similar temperature components as discussed in Sections \ref{sec_rd} and \ref{sec_n}. 
We thus applied $T_{d}$ = 60 K for the present source. 
The continuum brightness of SMM1 is measured to be 11.33 $\pm$ 0.05 mJy/beam for 1200 $\mu$m and 28.0 $\pm$ 0.2 mJy/beam for 870 $\mu$m (3$\sigma$ uncertainty). 
Based on the above assumption, we obtain $N_{\mathrm{H_2}}$ = 1.6 $\times$ 10$^{24}$ cm$^{-2}$ for 1200 $\mu$m and $N_{\mathrm{H_2}}$ = 1.2 $\times$ 10$^{24}$ cm$^{-2}$ for the 870 $\mu$m. 
The $N_{\mathrm{H_2}}$ value changes by a factor of up to 1.6 when the assumed $T_{d}$ is varied between 40 K and 90 K. 

Alternatively, a column density of molecular hydrogen can be determined by the model fitting of the observed spectral energy distribution (SED). 
The best-fit SED discussed in Section \ref{sec_disc_star} yields $A_V$ = 184 mag. 
We here use a standard value of $N_{\mathrm{H}}$/$E(B-V)$ = 5.8 $\times$ 10$^{21}$ cm$^{-2}$ mag$^{-1}$ \citep{Dra03} and a slightly high $A_{V}$/$E(B-V)$ ratio of 4 for dense clouds \citep{Whi01b}. 
Taking into account a factor of four lower metallicity, we obtain $N_{\mathrm{H_2}}$/$A_{V}$ = 2.9 $\times$ 10$^{21}$ cm$^{-2}$ mag$^{-1}$, where we assume that all the hydrogen atoms are in the form of H$_2$. 
Using this conversion factor, we obtain $N_{\mathrm{H_2}}$ = 5.3 $\times$ 10$^{23}$ cm$^{-2}$. 
This $N_{\mathrm{H_2}}$ is similar to the $N_{\mathrm{H_2}}$ derived from the aforementioned method assuming $T_{d}$ = 150 K. 
Such $T_{d}$ may be somewhat high as a typical dust temperature in the line of sight, but it is not very unrealistic value given the observed temperature range of molecular gas towards WB89-789 SMM1. 

In this paper, we use $N_{\mathrm{H_2}}$ = 1.1 $\times$ 10$^{24}$ cm$^{-2}$ as a representative value, which corresponds to the average of $N_{\mathrm{H_2}}$ derived by the dust continuum data and the SED fitting. 
This $N_{\mathrm{H_2}}$ corresponds to $A_V$ = 380 mag using the above conversion factor. 
Assuming the source diameter of 0.026 pc and the uniform spherical distribution of gas around a protostar, we estimate the gas number density to be $n_{\mathrm{H_2}}$ = 2.1 $\times$ 10$^7$ cm$^{-3}$, where the total gas mass of 13 M$_{\sun}$ is enclosed. 

\added{
Similarly, the mass for a 0.1 pc diameter region (i.e., a canonical size of dense cores) is estimated to be 75 M$_{\sun}$ with $T_{d}$ = 60 K, where Band 6 and Band 7 estimates are averaged. 
For the whole field shown in Figures \ref{images1}--\ref{images2}, which roughly corresponds to a 0.5 pc diameter region, the total mass is estimated to be 800--2500 M$_{\sun}$, where we assume $T_{d}$ = 20--10 K for extended dust emission. 
Note that this is a lower limit because the maximum recoverable scale of the present observations is 5$\farcs$4 (0.28 pc). 
}

\subsubsection{Fractional abundances and isotope abundance ratios} \label{sec_x} 
Fractional abundances with respect to H$_2$ are shown in Table \ref{tab_X}, which are calculated based on column densities estimated in Sections \ref{sec_rd}--\ref{sec_h2}. 
The fractional abundances normalized by the CH$_3$OH column density are also discussed in Sections \ref{sec_disc_molab}-\ref{sec_disc_molab2}, because of the non-negligible uncertainty associated with $N_{\mathrm{H_2}}$ (see Section \ref{sec_h2}). 

Abundances of HCO$^{+}$, HCN, SO, CS, OCS, and CH$_3$OH are estimated from their isotopologues, H$^{13}$CO$^{+}$, H$^{13}$CN, $^{34}$SO, C$^{34}$S, O$^{13}$CS, and $^{13}$CH$_3$OH. 
Detections of isotopologue species for SO, CS, OCS, and CH$_3$OH imply that the main species would be optically thick. 
Isotope abundance ratios of $^{12}$C/$^{13}$C = 150 and $^{32}$S/$^{33}$S = 35 are assumed, which are obtained by extrapolating the relationship between isotope ratios and galactocentric distances reported in \citet{Wil94} and \citet{Hum20} to $D_{GC}$ = 19 kpc.

Abundance ratios are derived for several rare isotopologues; we obtain CH$_2$DOH/CH$_3$OH = 0.011 $\pm$ 0.002, D$_2$CO/HDCO = 0.45  $\pm$ 0.10, $^{34}$SO/$^{33}$SO  = 5 $\pm$ 1, C$^{34}$S/C$^{33}$S = 2 $\pm$ 1, and $^{32}$SO$_2$/$^{34}$SO$_2$ = 20  $\pm$ 4. 
The $^{32}$SO$_2$/$^{34}$SO$_2$ ratio in WB89-789 SMM1 is similar to the solar $^{32}$S/$^{34}$S ratio \citep[22,][]{Wil94}, although we expect a slightly higher value in the outer Galaxy due to the $^{32}$S/$^{34}$S gradient in the Galaxy \citep{Chi96b, Hum20}. 
Astrophysical implication for the deuterated species are discussed in Section \ref{sec_disc_molab2}. 

The rotation diagram of CH$_3$CN is rather scattered. 
Although its isotopologue line is not detected, optical thickness might affect the column density estimate, as CH$_3$CN is often optically thick in hot cores \citep[e.g., ][]{Fue14}. 
To obtain a possible range of its column density, we use the rotation diagram of $^{12}$CH$_3$CN data to estimate a lower limit and the non-detection of the $^{13}$CH$_3$CN(19$_{0}$--18$_{0}$) line at 339.36630 GHz ($E_{u}$ = 163 K) for an upper limit. 

We have also repeated the analysis for the spectra extracted from a 0.1 pc (1$\farcs$93) diameter region at the hot core position, for the sake of comparison with LMC hot cores (see Section \ref{sec_disc_molab2}). 
Those abundances are also summarized in Table \ref{tab_X}. 
The abundances for a 0.1 pc area do not drastically vary from those for a 0.026 pc area. 
Molecules with compact spatial distribution (e.g., COMs) tend to decrease their abundances by a factor of $\sim$2--3 in the 0.1 pc data due to the beam dilution effect. 
In contrast, those with extended spatial distributions and intensity peaks outside the hot core region (H$^{13}$CO$^+$, CCH, CN, and NO) increases by a factor of $\sim$2 in the 0.1 pc data.

\begin{deluxetable}{ l c c c c}
\tablecaption{Estimated rotation temperatures, column densities, and source sizes \label{tab_N}}
\tabletypesize{\footnotesize} 
\tablehead{
\colhead{Molecule}   & \colhead{$T$$_{rot}$}   &       \colhead{$N$(X)}            &  \colhead{$N$(X) non-LTE}  & \colhead{Size}  \\
\colhead{ }                & \colhead{(K)}                 &        \colhead{(cm$^{-2}$)}    & \colhead{(cm$^{-2}$)}           &   \colhead{($\arcsec$)}
}
\startdata 
H$_2$                                                &   \nodata                           &   1.1 $\times$ 10$^{24}$   & \nodata   & 0.85\tablenotemark{c}  \\
\tableline
H$^{13}$CO$^+$                              &   35                                   &   (7.0 $\pm$ 0.1) $\times$ 10$^{12}$   &  (7.6 $\pm$ 0.9) $\times$ 10$^{12}$  & $>$1.5\tablenotemark{d}    \\
HC$^{18}$O$^+$                              &   35                                   &   (5.8 $\pm$ 0.9) $\times$ 10$^{11}$   &   (5.7 $\pm$ 0.6) $\times$ 10$^{11}$  & 1.18\tablenotemark{d}   \\
CCH                                                  &   35                                    &   (2.7 $\pm$ 0.1) $\times$ 10$^{14}$   & \nodata  &  $>$2\tablenotemark{d}  \\
c-C$_3$H$_2$                                 &   35                                    &   (9.5 $\pm$ 2.2) $\times$ 10$^{13}$   &  (8.2 $\pm$ 0.9) $\times$ 10$^{13}$\tablenotemark{a}  &  $>$1\tablenotemark{d}    \\
H$_2$CO                                          &   39                                    &   (1.1 $\pm$ 0.1) $\times$ 10$^{14}$   &  (1.3 $\pm$ 0.1) $\times$ 10$^{14}$\tablenotemark{a}  &  $>$1.5\tablenotemark{d}   \\
HDCO                                               &   39                                    &   (5.1 $\pm$ 0.3) $\times$ 10$^{13}$   &  \nodata &  $>$1\tablenotemark{d}  \\
D$_2$CO                                         &   \textit{39$^{+6}_{-5}$}                 &   \textit{(2.3 $\pm$ 0.5) $\times$ 10$^{13}$}   &  \nodata\nodata &  $>$1\tablenotemark{d}  \\
 CN                                                  &   35                                   &   (3.3 $\pm$ 0.2) $\times$ 10$^{14}$  &  (2.5 $\pm$ 0.3) $\times$ 10$^{14}$  &  $>$2\tablenotemark{d}   \\ 
H$^{13}$CN                                    &   35                                   &   (1.2 $\pm$ 0.1) $\times$ 10$^{13}$   &  (1.1 $\pm$ 0.1) $\times$ 10$^{13}$   &  0.92\tablenotemark{d}   \\
HC$^{15}$N                                    &   35                                   &   (6.3 $\pm$ 0.2) $\times$ 10$^{12}$   &   (5.8 $\pm$ 0.6) $\times$ 10$^{12}$  &   0.75\tablenotemark{d}  \\
HC$_3$N                                        &   200                                  &   (2.7 $\pm$ 0.3) $\times$ 10$^{13}$   &  (2.1 $\pm$ 0.2) $\times$ 10$^{13}$  &   0.65   \\
 NO                                                 &   35                                    &  (9.0 $\pm$ 2.5) $\times$ 10$^{14}$    &  (8.9 $\pm$ 0.9) $\times$ 10$^{14}$    &  $>$1.5\tablenotemark{d}   \\ 
 HNCO                                            &   \textit{237$^{+17}_{-15}$}          &   \textit{(3.0 $\pm$ 0.2) $\times$ 10$^{14}$}   &  \nodata   &  0.54  \\  
 CH$_3$CN                                    &   \textit{279$^{+12}_{-11}$}           &   \textit{(1.8 $\pm$ 0.1) $\times$ 10$^{14}$}   &  (8.6 $\pm$ 0.8) $\times$ 10$^{13}$  &   0.51    \\
$^{13}$CH$_3$CN                        &   200                                  &   $<$5 $\times$ 10$^{12}$                    &  \nodata  &   \nodata  \\
 C$_2$H$_5$CN                            &   \textit{130$^{+20}_{-15}$}           &   \textit{(6.3 $\pm$ 1.7) $\times$ 10$^{13}$}    &  \nodata  &  0.52   \\
 NH$_2$CO                                    &   \textit{140$^{+8}_{-7}$}              &    \textit{(4.2 $\pm$ 0.7) $\times$ 10$^{13}$}   & \nodata   &   0.56  \\
 SiO                                                 &   35                                    &   (2.5 $\pm$ 0.2) $\times$ 10$^{12}$   & (2.5 $\pm$ 0.3) $\times$ 10$^{12}$  &  0.65    \\
 CS                                                  &   36                                   &   (1.5 $\pm$ 0.2) $\times$ 10$^{14}$    &  (2.0 $\pm$ 0.3) $\times$ 10$^{14}$   &   $>$1.5  \\
 C$^{34}$S                                      &   36                                   &   (3.1 $\pm$ 0.1) $\times$ 10$^{13}$   &  \nodata  &  0.70  \\
 C$^{33}$S                                      &   \textit{36$^{+4}_{-3}$}               &   \textit{(1.5 $\pm$ 0.2) $\times$ 10$^{13}$}   &  \nodata  &   0.61   \\
 OCS                                               &   \textit{106$^{+6}_{-5}$}            &   \textit{(6.5 $\pm$ 0.5) $\times$ 10$^{14}$}    &   (6.4 $\pm$ 0.7) $\times$ 10$^{14}$  &  0.55   \\ 
 $^{13}$OCS                                   &   200                                 &   (8.7 $\pm$ 2.4) $\times$ 10$^{13}$   &   \nodata  &   0.45  \\ 
 H$_2$CS                                       &   \textit{43$^{+3}_{-2}$}                &   \textit{(1.5 $\pm$ 0.1) $\times$ 10$^{14}$}   &   (1.4 $\pm$ 0.2) $\times$ 10$^{14}$\tablenotemark{a}  &   0.62  \\
 SO                                                 &    \textit{35$^{+1}_{-1}$}               &   \textit{(4.0 $\pm$ 0.3) $\times$ 10$^{14}$}    &   (4.5 $\pm$ 0.5) $\times$ 10$^{14}$   &   0.70\tablenotemark{d}   \\
$^{34}$SO                                      &    35                                   &   (5.9 $\pm$ 0.1) $\times$ 10$^{13}$    &  \nodata  &    0.66  \\
 $^{33}$SO                                     &    35                                   &   (1.1 $\pm$ 0.1) $\times$ 10$^{13}$    &  \nodata  &    0.53 \\
 SO$_2$                                         &    \textit{166$^{+5}_{-5}$}             &   \textit{(1.2 $\pm$ 0.1) $\times$ 10$^{15}$}    &   \nodata  &   0.53    \\
 $^{34}$SO$_2$                             &    166                                 &   (5.9 $\pm$ 0.9) $\times$ 10$^{13}$    &  \nodata   &   0.51   \\
 CH$_3$SH                                    &    200                                 &   $<$3 $\times$ 10$^{14}$                    &  \nodata   &   \nodata    \\
HDO                                                 &   \textit{217$^{+14}_{-12}$}          &   \textit{(2.2 $\pm$ 0.2) $\times$ 10$^{15}$}    &  \nodata  &  0.52   \\
CH$_3$OH                                       &   \textit{245$^{+4}_{-4}$}              &   \textit{(1.9 $\pm$ 0.1) $\times$ 10$^{16}$}    & (2.6 $\pm$ 0.1) $\times$ 10$^{16}$\tablenotemark{b} &   0.51   \\
$^{13}$CH$_3$OH                          &   \textit{181$^{+10}_{-9}$}            &   \textit{(2.8 $\pm$ 0.2) $\times$ 10$^{15}$}     &  \nodata   &   0.46 \\
CH$_2$DOH                                   &   \textit{155$^{+18}_{-15}$}           &   \textit{(4.6 $\pm$ 0.3) $\times$ 10$^{15}$}   &   \nodata & 0.52  \\
HCOOCH$_3$                                &   \textit{181$^{+6}_{-5}$}               &   \textit{(8.6 $\pm$ 0.4) $\times$ 10$^{15}$}    &  \nodata & 0.51  \\
CH$_3$OCH$_3$                           &   \textit{137$^{+5}_{-4}$}               &   \textit{(2.6 $\pm$ 0.1) $\times$ 10$^{15}$}    &  \nodata &  0.52 \\  
C$_2$H$_5$OH                              &   \textit{136$^{+14}_{-12}$}          &   \textit{(9.6 $\pm$ 1.3) $\times$ 10$^{14}$}   & \nodata   &  0.50 \\
CH$_3$CHO                                   &   \textit{192$^{+52}_{-34}$}          &   \textit{(6.4 $\pm$ 0.8) $\times$ 10$^{14}$}    & \nodata  &  0.49  \\
\textit{trans}-HCOOH                      &   \textit{71$^{+11}_{-9}$}               &   \textit{(2.7 $\pm$ 0.6) $\times$ 10$^{14}$}   &  \nodata  & 0.58  \\
\textit{cis}-HCOOH                          &   \textit{69$^{+50}_{-21}$}            &   \textit{(2.4 $\pm$ 1.2) $\times$ 10$^{13}$}    & \nodata   & 0.49  \\
H$_2$CCO                                      &  \textit{92$^{+14}_{-11}$}           &   \textit{(1.0 $\pm$ 0.2) $\times$ 10$^{14}$}    &  \nodata  & 0.55  \\
 c-C$_2$H$_4$O                            &   200                                 &   (8.9 $\pm$ 2.0) $\times$ 10$^{13}$     &  \nodata &  0.47  \\
\enddata
\tablecomments{
\added{For $T$$_{rot}$ and $N$(X), those derived by rotation diagrams are shown in italics. }
Uncertainties and upper limits are of the 2 $\sigma$ level and do not include systematic errors due to adopted spectroscopic constants. 
See Sections \ref{sec_rd}-\ref{sec_h2} and \ref{sec_disc_dist} for details. 
}
\tablenotetext{a}{Assuming ortho/para ratio of three. }
\tablenotetext{b}{Assuming E-CH$_3$OH/A-CH$_3$OH ratio of unity \citep{Wir11}. }
\tablenotetext{c}{Size of continuum emission. }
\tablenotetext{d}{Associated with extended component. }
\end{deluxetable}

\begin{deluxetable}{ l c c } 
\tablecaption{Estimated fractional abundances \label{tab_X}} 
\tabletypesize{\small} 
\tablehead{
\colhead{Molecule}                         & \multicolumn{2}{c}{$N$(X)/$N_{\mathrm{H_2}}$}     \\
\colhead{}                                       & \colhead{0.026 pc area}         & \colhead{0.1 pc area}       
}
\startdata 
   HCO$^+$\tablenotemark{a}         & (9.5 $\pm$ 3.2) $\times$ 10$^{-10}$      &  (1.5 $\pm$ 0.3) $\times$ 10$^{-9}$     \\
   H$_2$CO                                    &  (1.0 $\pm$ 0.3) $\times$ 10$^{-10}$    &  (1.2 $\pm$ 0.1) $\times$ 10$^{-10}$  \\
   HDCO                                          &   (4.7 $\pm$ 1.3) $\times$ 10$^{-11}$   &  (3.9 $\pm$ 0.2) $\times$ 10$^{-11}$      \\
   D$_2$CO                                    &  (2.1 $\pm$ 0.7) $\times$ 10$^{-11}$    &  (2.0 $\pm$ 0.3) $\times$ 10$^{-11}$   \\
   C$_2$H                                     &  (2.5 $\pm$ 0.7) $\times$ 10$^{-10}$     &   (5.8 $\pm$ 1.2) $\times$ 10$^{-10}$      \\
   c-C$_3$H$_2$                          &   (8.6 $\pm$ 3.1) $\times$ 10$^{-11}$    &    (5.9 $\pm$ 1.2) $\times$ 10$^{-11}$  \\
   CN                                            &   (3.0 $\pm$ 0.8) $\times$ 10$^{-10}$     &    (6.6 $\pm$ 1.3) $\times$ 10$^{-10}$    \\   
   HCN\tablenotemark{a}            &   (1.7 $\pm$ 0.6) $\times$ 10$^{-9}$      &    (1.2 $\pm$ 0.3) $\times$ 10$^{-9}$   \\
   HC$_3$N                                 &   (2.5 $\pm$ 0.7) $\times$ 10$^{-11}$     &   (1.4 $\pm$ 0.1) $\times$ 10$^{-11}$    \\
   NO                                            &   (8.1 $\pm$ 3.2) $\times$ 10$^{-10}$     &  (1.6 $\pm$ 0.1) $\times$ 10$^{-9}$   \\   
   HNCO                                       &   (2.7 $\pm$ 0.8) $\times$ 10$^{-10}$     &  (7.1 $\pm$ 0.6) $\times$ 10$^{-11}$    \\
   CH$_3$CN\tablenotemark{b}       &  (4.2 $\pm$ 2.7) $\times$ 10$^{-10}$     &   (3.7 $\pm$ 2.8) $\times$ 10$^{-10}$       \\
   C$_2$H$_5$CN                                 &  (5.8 $\pm$ 2.2) $\times$ 10$^{-11}$  &    (2.4 $\pm$ 0.9) $\times$ 10$^{-11}$     \\
   NH$_2$CHO                                  &   (3.8 $\pm$ 1.2) $\times$ 10$^{-11}$     &  (1.8 $\pm$ 0.1) $\times$ 10$^{-11}$    \\
   SiO                                      &  (2.2 $\pm$ 0.6) $\times$ 10$^{-12}$        &   (1.2 $\pm$ 0.1) $\times$ 10$^{-12}$      \\
   CS\tablenotemark{c}           &  (9.7 $\pm$ 3.3) $\times$ 10$^{-10}$         &   (6.4 $\pm$ 1.3) $\times$ 10$^{-10}$    \\
   SO\tablenotemark{c}            & (1.9 $\pm$ 0.5) $\times$ 10$^{-9}$         &   (1.3 $\pm$ 0.3) $\times$ 10$^{-9}$     \\
   OCS\tablenotemark{a}         &  (1.2 $\pm$ 0.5) $\times$ 10$^{-8}$     &    (4.1 $\pm$ 1.4) $\times$ 10$^{-9}$    \\
   H$_2$CS                             &  (1.4 $\pm$ 0.4) $\times$ 10$^{-10}$      &    (9.0 $\pm$ 1.0) $\times$ 10$^{-11}$    \\
   SO$_2$                             &   (1.1 $\pm$ 0.3) $\times$ 10$^{-9}$         &   (2.9 $\pm$ 0.1) $\times$ 10$^{-10}$    \\
   CH$_3$SH                                   & $<$3 $\times$ 10$^{-10}$               &   $<$2 $\times$ 10$^{-10}$    \\
   HDO                                      &  (2.0 $\pm$ 0.6) $\times$ 10$^{-9}$       &   (7.7 $\pm$ 0.9) $\times$ 10$^{-10}$   \\
   CH$_3$OH\tablenotemark{a}      &  (3.8 $\pm$ 1.3) $\times$ 10$^{-7}$           & (1.7 $\pm$ 0.3) $\times$ 10$^{-7}$     \\
   CH$_2$DOH                                  &  (4.2 $\pm$ 1.2) $\times$ 10$^{-9}$         &  (1.5 $\pm$ 0.2) $\times$ 10$^{-9}$      \\
   HCOOCH$_3$                                 &  (7.8 $\pm$ 2.2) $\times$ 10$^{-9}$       &  (3.0 $\pm$ 0.2) $\times$ 10$^{-9}$    \\
   CH$_3$OCH$_3$                                & (2.3 $\pm$ 0.6) $\times$ 10$^{-9}$    &  (1.0 $\pm$ 0.1) $\times$ 10$^{-9}$  \\   
   C$_2$H$_5$OH                                 &   (8.7 $\pm$ 2.7) $\times$ 10$^{-10}$  &  (3.3 $\pm$ 0.8) $\times$ 10$^{-10}$         \\
   CH$_3$CHO                                  &  (5.8 $\pm$ 1.8) $\times$ 10$^{-10}$        &  (2.1 $\pm$ 0.4) $\times$ 10$^{-10}$    \\
   HCOOH\tablenotemark{d}       &  (2.7 $\pm$ 1.0) $\times$ 10$^{-10}$              &  (1.2 $\pm$ 0.4) $\times$ 10$^{-10}$     \\
   H$_2$CCO                                   & (9.2 $\pm$ 3.0) $\times$ 10$^{-11}$          &   (3.7 $\pm$ 0.9) $\times$ 10$^{-11}$     \\
   c-C$_2$H$_4$O                                &   (8.1 $\pm$ 2.8) $\times$ 10$^{-11}$   &   (5.9 $\pm$ 1.2) $\times$ 10$^{-11}$   \\
\enddata
\tablecomments{
Uncertainties and upper limits are of the 2$\sigma$ level. 
Column densities of molecules for a 0.026 pc area are summarized in Table \ref{tab_N}. 
An empirical uncertainty of 30 $\%$ is assumed for $N_{\mathrm{H_2}}$. 
}
\tablenotetext{a}{Estimated from $^{13}$C isotopologue with $^{12}$C/$^{13}$C = 150 }
\tablenotetext{b}{Rotation diagram analysis of CH$_3$CN is used to derive a lower limit and the non-detection of $^{13}$CH$_3$CN for an upper limit. }
\tablenotetext{c}{Estimated from $^{34}$S isotopologue with $^{32}$S/$^{34}$S = 35 }
\tablenotetext{d}{Sum of $trans-$ and $cis-$species. }
\end{deluxetable}

\begin{figure}[tpbh!]
\begin{center}
\includegraphics[width=8.5cm]{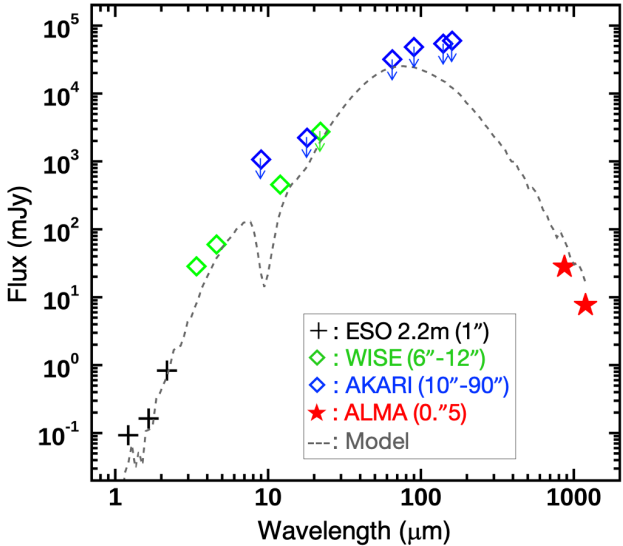}
\caption{
The SED of WB89-789 SMM1. 
The plotted data are obtained by the ESO 2.2 m telescope \citep[pluses, black; ][]{Bra07}, the WISE all-sky survey \citep[open diamonds, light green; ][]{Wri10}, \textit{AKARI} FIS all-sky survey \citep[open diamonds, blue; ][]{Yam10}, and ALMA (filled star, red, this work). 
The angular resolution of each data is indicated in brackets. 
The gray dashed line indicates the best-fitted SED with the model of \citet{Rob07}. 
}
\label{sed}
\end{center}
\end{figure}

\section{Discussion} \label{sec_disc} 
\subsection{Hot molecular core and protostar associated with WB89-789 SMM1} \label{sec_disc_star} 
The nature of WB89-789 SMM1 is characterized as 
(i) the compact distribution of warm gas ($\sim$0.03 pc, see Section \ref{sec_disc_dist}), 
(ii) the high gas temperature that can trigger the ice sublimation ($\geq$100 K, Section \ref{sec_rd}), 
(iii) the high density (2 $\times$ 10$^7$ cm$^{-3}$, Section \ref{sec_h2}), 
(iv) the association with a luminous protostar (see below), 
and (v) the presence of chemically rich molecular gas. 
Those properties suggest that the source is associated with a hot molecular core. 

Figure \ref{sed} shows a SED of SMM1, where the data are collected from available databases and literatures \citep{Bra07, Wri10, Yam10}. 
The bolometric luminosity of the source is estimated to be 8.4 $\times$ 10$^3$ L$_{\sun}$ based on the SED fitting with the model of \citet{Rob07}. 
This luminosity is equivalent to a stellar mass of about 10 M$_{\sun}$ according to the mass-luminosity relationship of zero age main sequence (ZAMS) stars \citep{ZY07}. 

Note that far-infrared data, which is important for the luminosity determination of embedded sources, is insufficient for SMM1. 
Only upper limits are provided due to the low angular resolution of available \textit{AKARI} FIS all-sky survey data. 
Thus the derived luminosity (and therefore mass) may be lower than the current estimate. 
Future high spatial resolution infrared observations in those missing wavelengths are highly required. 

Alternatively, we can estimate the luminosity of SMM1 by scaling the luminosity of a low-metallicity LMC hot core, ST16, whose SED is well determined based on a comprehensive infrared \replaced{data set}{dataset} from 1 to 1200 $\mu$m \citep{ST20}. 
This LMC hot core has a total luminosity of 3.1 $\times$ 10$^5$ L$_{\sun}$ and a $K_s$-band magnitude ([$K_s$]) of 13.4 mag at 50 kpc, while SMM1 has [$K_s$] = 14.7 mag at 10.7 kpc. 
Scaling the luminosity of ST16 with the distance and $K_s$-band magnitude, we obtain 4.3 $\times$ 10$^3$ L$_{\sun}$ for SMM1, which is a factor of two lower than the estimate by the SED fitting. 
In either case, present estimates suggest that the luminosity of SMM1 would correspond to the lower-end of high-mass ZAMS or upper-end of intermediate-mass ZAMS. 

\begin{figure}[tp!]
\begin{center}
\includegraphics[width=8.7cm]{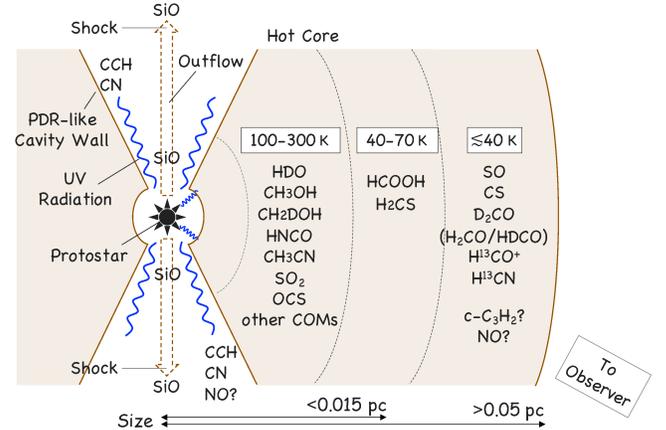}
\caption{
Schematic illustration of the molecular gas distribution and the temperature structure in WB89-789 SMM1. 
}
\label{schematic}
\end{center}
\end{figure}
%
\subsection{Distribution of molecular line emission and dust continuum} \label{sec_disc_dist} 
The observed emission lines and continuum show different spatial distributions depending on species. 
Those distributions have important clues to understand their origins. 
A schematic illustration of the temperature structure and molecular gas distribution in WB89-789 SMM1 are shown in Figure \ref{schematic} based on the discussion in this section. 

We have estimated the spatial extent of observed emission by fitting a two-dimensional Gaussian to the continuum center (Table \ref{tab_N}). 
Compact distributions (FWHM = 0$\farcs$5--0$\farcs$6, 0.026--0.031 pc), that is comparable with the beam size, are seen in HDO, COMs, CH$_3$CN, HNCO, OCS, and high-excitation SO$_2$ lines. 
HC$_3$N is slightly extended (FWHM = 0$\farcs$65). 
They are concentrated at the hot core position, suggesting that they are originated from a warm region where ice mantles are sublimated. 

SO, $^{34}$SO, $^{33}$SO, and low-excitation SO$_2$ show relatively compact distributions (FWHM = 0$\farcs$5--0$\farcs$7, 0.026--0.036 pc) at the hot core position, but also show a secondary peak at the south side of the hot core. 
This secondary peak coincides with the peak of the NO emission. 
Other sulfur-bearing species such as C$^{34}$S, C$^{33}$S, and H$_2$CS show compact distributions \replaced{(FWHM = 0$\farcs$6--0$\farcs$0.7 (0.031--0.052 pc)}{(FWHM = 0$\farcs$6--0$\farcs$0.7, 0.031--0.052 pc)} centered at the hot core. 

A characteristic distribution that is symmetric to the hot core position is seen in SiO. 
It shows a compact emission (FWHM = 0$\farcs$65) at the hot core center, but also shows other peaks at the north-east and south-west sides of the hot core. 
Those secondary peaks are slightly elongated. 
SiO is a well-known shock tracer. 
The observed structure would be originated from the shocked gas produced by bipolar protostellar outflows. 
A driving source of the outflows would be a protostar embedded in a hot core, since the distribution of SiO is symmetric to the hot core position. 

Even extended distributions (FWHM $>$ 1$\farcs$0) are seen in CN, CCH, H$^{13}$CO$^{+}$, HC$^{18}$O$^{+}$, H$^{13}$CN, HC$^{15}$N, NO, CS, H$_2$CO, and HDCO, D$_2$CO, and low-excitation CH$_3$OH. 
Gas-phase reactions and non-thermal desorption of icy species would have non-negligible contribution to the formation of those species, because they are widely distributed beyond the hot core. 
We note that dust continuum, H$^{13}$CN, HC$^{15}$N have a moderately sharp peak (FWHM $<$ 1$\farcs$0) at the hot core position in addition to the extended component. 
c-C$_3$H$_2$ shows a patchy distribution, whose secondary peak at the south-west of the hot core does not \replaced{coincides}{coincide} with those of other species. 

Molecular radicals (CN, CCH, and NO) do not have their emission peak at the hot core position. 
This would suggest that the chemistry outside the hot core region largely contributes to their production. 
CN and CCH are known to be abundant in photodissociation regions (PDRs), because atomic carbon is efficiently provided by photodissociation of CO under moderate UV fields \citep[e.g., ][]{Fue93, Ste95, Jan95, Rod98, Pet17}. 
In the present source, their emission shows a similar spatial distribution. 
A similar distribution between CN and CCH has been also observed in a LMC hot core \citep{ST20}; they argue that CN and CCH would trace PDR-like outflow cavity structures that are irradiated by the UV light from a protostar associated with a hot core. 
We speculate that this is also the case for WB89-789 SMM1. 

\begin{figure}[tp!]
\begin{center}
\includegraphics[width=8.5cm]{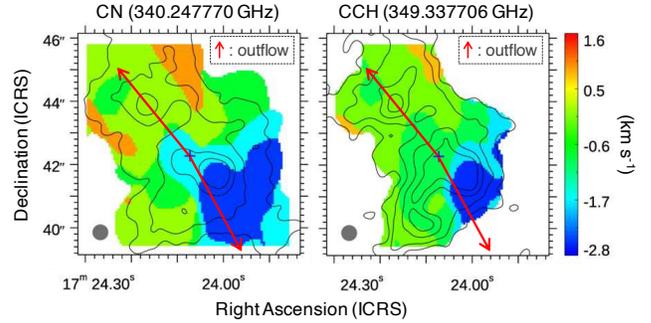}
\caption{
Velocity maps (moment 1) of CN and CCH lines. 
The color scale indicates the offset velocity relative to the systemic velocity of 34.5 km s$^{-1}$. 
A possible direction of outflows expected from the distributions of SiO is shown by the red arrows.  
Contours represent the integrated intensity distribution and the contour levels are 8$\%$, 20$\%$, 40$\%$, and 60$\%$ of the peak value. 
Low signal-to-noise regions (S/N $<$5) are masked. 
The blue cross represents the 1.2 mm continuum center. }
\label{Mom1}
\end{center}
\end{figure}

Figure \ref{Mom1} shows velocity maps (moment 1) of CN and CCH lines. 
CN and CCH emission are elongated in the south-west direction from the hot core (see also Figure \ref{images1}). 
The figure also shows a possible direction of protostellar outflows expected from the spatial distribution of SiO. 
The elongated directions of CN and CCH coincide with the inferred direction of outflows. 
In addition, the elongated south-west parts of CN and CCH are blue-shifted by $\sim$1--2 km s$^{-1}$ compared to the hot core position. 
This may be due to outflow gas motion, although CN and CCH would trace an outflow cavity wall rather than outflow gas itself. 
Actually the observed velocity shift is smaller than a typical value of high-velocity wing components in massive protostellar outflows \citep[$\geq$ 5 km s$^{-1}$, e.g., ][]{Beu02, Mau15}. 
\added{We note that a clear velocity structure is not seen in the SiO velocity map, except for the position of another embedded protostar discussed in Section \ref{sec_disc_SiO}. }
Future observations of optically-thick outflow tracers such as CO are necessary to confirm the presence of high-velocity gas associated with protostellar outflows. 

\subsection{Molecular abundances: Comparison with Galactic hot cores} \label{sec_disc_molab} 
\begin{figure*}[tpb!]
\begin{center}
\includegraphics[width=17.0cm]{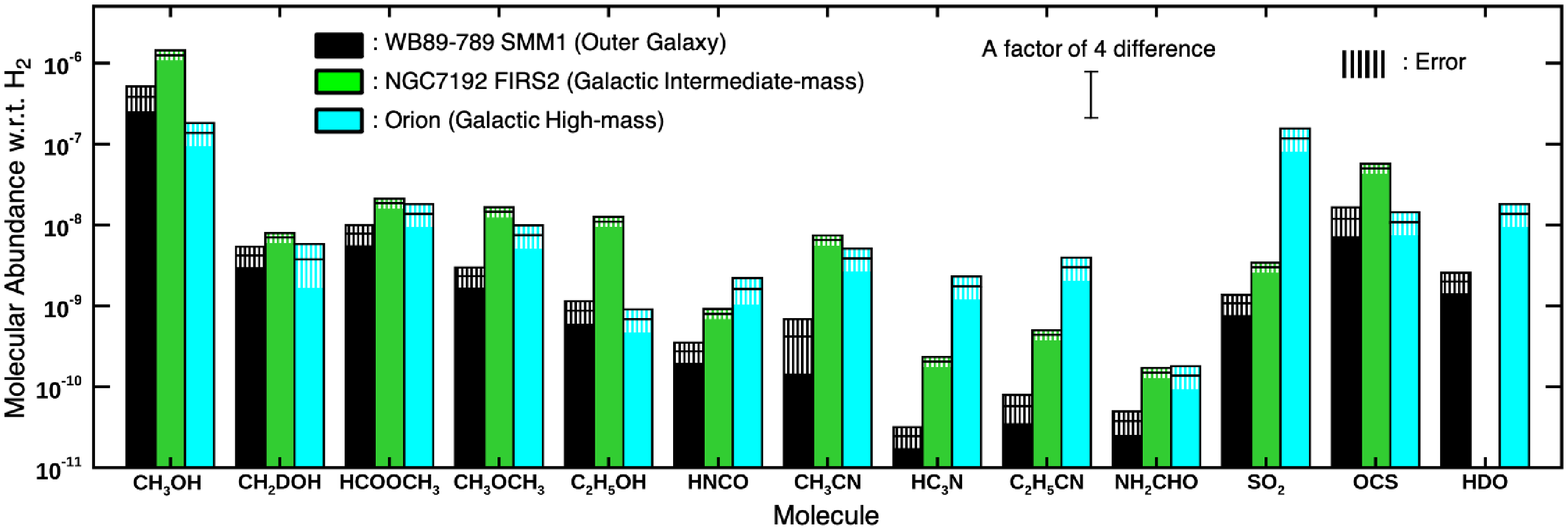}
\caption{
Comparison of molecular abundances between an outer Galactic hot core (black, WB89-789 SMM1), an intermediate-mass hot core (green, NGC7192 FIRS2), and a high-mass hot core (cyan, Orion). 
An abundance difference by a factor of four is indicated by the black solid line with hats. 
The area with thin vertical lines indicate the error bar. 
No data is available for HDO in NGC7192 FIRS2. 
See Section \ref{sec_disc_molab} for details. 
}
\label{abu1}
\end{center}
\end{figure*}

Figure \ref{abu1} shows a comparison of molecular abundances between WB89-789 SMM1 and other known Galactic hot cores. 
The data for an intermediate-mass hot core, NGC7192 FIRS2, is adopted from \citet{Fue14}. 
The abundances are based on the 220 GHz region observations for a 0.009 pc diameter area centered at the hot core. 
The luminosity of NGC7192 FIRS2 ($\sim$500 L$_{\sun}$) corresponds to that of a 5 M$_{\sun}$ ZAMS. 
The data for a high-mass source, the Orion hot core, is adopted from \citet{Sut95}, which is based on the 340 GHz region observations for a 0.027 pc diameter area at the hot core. 
The abundance of HNCO is taken from \citet{Sch97}. 

The molecular abundances in WB89-789 SMM1 is generally lower than those of inner Galactic counterparts. 
The degree of the abundance decrease is roughly consistent with the lower metallicity of the WB89-789 region as indicated by the scale bar in Figure \ref{abu1}. 
Particularly, SMM1 and the intermediate-mass hot core NGC7192 FIRS2 show similar molecular abundances after taking into account the four times lower metallicity of the former source.
For the comparison with Orion, it seems that HC$_3$N, C$_2$H$_5$CN, and SO$_2$ are significantly less abundant in SMM1 even taking into account the lower metallicity, while CH$_3$OH is overabundant in SMM1 despite the low metallicity. 

\begin{figure*}[tpb!]
\begin{center}
\includegraphics[width=18.0cm]{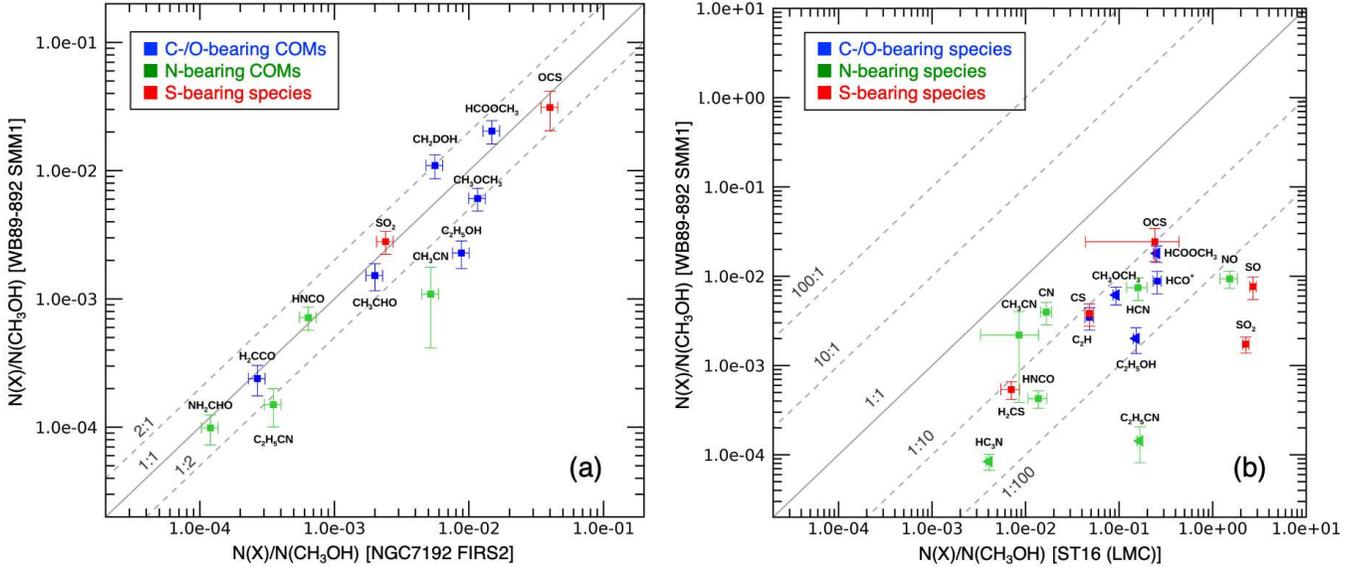}
\caption{
Comparison of molecular abundances normalized by the CH$_3$OH column density for (a) WB89-789 SMM1 vs. NGC7192 FIRS2 and (b) WB89-789 SMM1 vs. ST16 (LMC). 
Carbon- and oxygen-bearing species are shown by the blue squares, nitrogen-bearing species in green, and sulfur-bearing species in red. 
The dotted lines in the panel (a) represent an abundance ratio of 2:1 and 1:2 for WB89-789 SMM1 : NGC7192 FIRS2, while the solid line represent that of 1:1. 
Similarly, the dotted lines in the panel (b) represent a ratio of 100:1, 10:1, 1:10, and 1:100 for WB89-789 SMM1:ST16, while 1:1 for the solid line. 
The leftward triangles in the panel (b) indicate the upper limit for ST16. 
See Section \ref{sec_disc_molab} for details. 
}
\label{abu2}
\end{center}
\end{figure*}

To further focus on chemical complexity at low metallicity, Figure \ref{abu2} shows a comparison of fractional abundances of COMs normalized by the CH$_3$OH column density for WB89-789 SMM1 and NGC7192 FIRS2. 
Such a comparison is useful for investigating chemistry of organic molecules in warm and dense gas around protostars \citep{Her09,Dro19}, because CH$_3$OH is believed to be a parental molecule for the formation of even larger COMs \citep[e.g.,][]{NM04,Gar06}. 
In addition, CH$_3$OH is a product of grain surface reaction, thus warm CH$_3$OH gas mainly arise from a high-temperature region, where ices are sublimated and characteristic hot core chemistry proceeds. 
Furthermore, the normalization by CH$_3$OH can cancel the metallicity effect in the abundance comparison. 

The $N$(X)/$N$(CH$_3$OH) ratios are remarkably similar between WB89-789 SMM1 and NGC7192 FIRS2 as shown in Figure \ref{abu2} (a). 
The ratios of SMM1 coincide with those of NGC7192 FIRS2 within a factor of 2 for the most molecular species. 
The correlation coefficient is calculated to be 0.94, while it is 0.96 if CH$_3$CN is excluded. 
It seems that CH$_3$CN deviates from the overall trend, although the uncertainty is large due to the opacity effect (see \ref{sec_x}). 
C$_2$H$_5$OH also slightly deviates from the trend. 
The reason for their behavior is still unclear, but it may be related to the formation pathway of those molecules. 

The above two comparisons suggest that chemical compositions of the hot core in the extreme outer Galaxy scale with the metallicity. 
In the WB89-789 region, the metallicity is expected to be four times lower compared to the solar neighborhood. 
The observed abundances of COMs in the SMM1 hot core is lower than the other Galactic hot cores, but the decrease is proportional to this metallicity. 
Furthermore, similar $N$(COMs)/$N$(CH$_3$OH) ratios suggest that CH$_3$OH is an important parental species for the formation of larger COMs in a hot core, as suggested by aforementioned theoretical studies. 

CH$_3$OH ice is believed to form on grain surfaces and several formation processes are proposed by laboratory experiments; i.e., hydrogenation of CO, ultraviolet photolysis and radiolysis of ice mixtures \citep[e.g.,][]{Hud99,Wat07}. 
It is known that CH$_3$OH is already formed in quiescent prestellar cores before star formation occurs \citep{Boo11}. 
Solid CH$_3$OH will chemically evolve to larger COMs by a combination of photolysis, radiolysis, and grain heating during the warm-up phase that leads to the formation of a hot core \citep{Gar06}. 
High-temperature gas-phase chemistry of sublimated CH$_3$OH would also contribute to the COMs formation \citep{NM04,Taq16}. 
The present results suggest that various COMs can form even in a low-metallicity environment, if their parental molecule, CH$_3$OH, is efficiently produced in a star-forming core. 
\added{The detection of a chemically-rich star-forming core in the extreme outer Galaxy has an impact on the understanding of the occurrence of the chemical complexity in a primordial environment of the early phase of the Galaxy formation. }
We here note that observations of ice mantle compositions are not reported for the outer Galaxy so far . 
Future infrared observations of ice absorption bands towards embedded sources in the outer Galaxy are important.

\begin{figure*}[tbp!]
\begin{center}
\includegraphics[width=16.5cm]{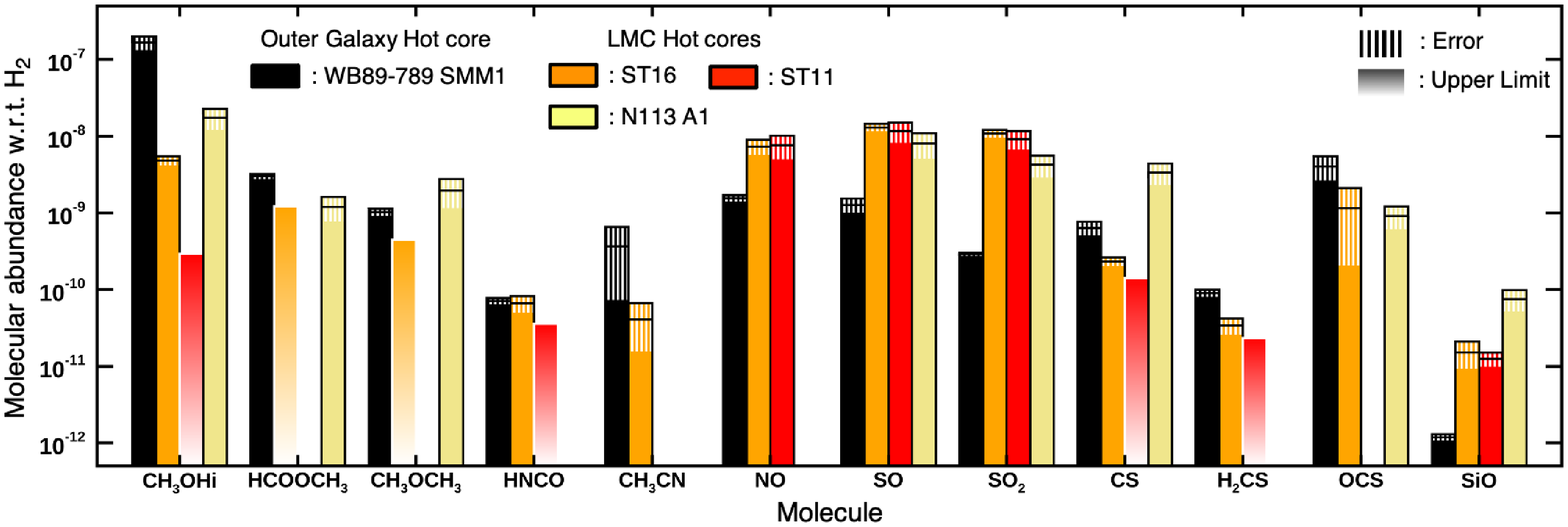}
\caption{
Comparison of molecular abundances between an outer Galactic hot core, WB89-789 SMM1 (black), and three LMC hot cores, ST11 (red), ST16 (orange), and N113 A1 (light yellow). 
Abundances of SMM1 are calculated for a 0.1 pc diameter region. 
The area with thin vertical lines indicate the error bar. 
The bar with a color gradient indicate an upper limit. 
The absence of bars indicate the lack of available data. 
See Section \ref{sec_disc_molab2} for details. 
}
\label{abu3}
\end{center}
\end{figure*}

\subsection{Molecular abundances: Comparison with LMC hot cores} \label{sec_disc_molab2} 
It is still unknown if the observed simply-metallicity-scaled chemistry of COMs in the WB89-789 SMM1 hot core is common in other hot core sources in the outer Galaxy. 
A comparison of the present data with those of hot cores in the LMC would provide a hint for understanding the universality of low-metallicity hot core chemistry. 
The metallicity of the LMC is reported to be lower than the solar value by a factor of two to three \citep[e.g.,][]{Duf82, Wes90, Rus92, Cho16}, which is in common with the outer Galaxy. 

Figure \ref{abu3} shows a comparison of molecular abundances between WB89-789 SMM1 and three LMC hot cores. 
The plotted molecular column densities for LMC hot cores are adopted from \citet{ST16} for ST11, \citet{ST20} for ST16, and \citet{Sew18} fro N113 A1. 
Another LMC hot core in \citet{Sew18}, N113 B3, have similar molecular abundances with those of N113 A1. 
The $N_{\mathrm{H_2}}$ value of ST11 and N113 A1 is re-estimated using the same dust opacity data and dust temperature ($T_{d}$ = 60 K) as in this work; 
We obtained $N_{\mathrm{H_2}}$ =  1.2 $\times$ 10$^{24}$ cm$^{-2}$ for ST11 and $N_{\mathrm{H_2}}$ =  9.2 $\times$ 10$^{23}$ cm$^{-2}$ for N113 A1. 
The dust temperature assumed in ST16 is 60 K as described in Section \ref{sec_h2}. 
Molecular column densities are estimated for circular/elliptical regions of 0.12 $\times$ 0.12 pc, 0.10 $\times$ 0.10 pc, and 0.21 $\times$ 0.13 pc for ST11, ST16, and N113 A1, respectively. 
For a fair comparison, we have re-calculated $N_{\mathrm{H_2}}$ and molecular column densities of SMM1 for a 0.1 pc (1$\farcs$93) diameter region. 
Those abundances are plotted in Figure \ref{abu3} and summarized in Table \ref{tab_X}. 

The chemical composition of the outer Galaxy hot core does not resemble those of LMC hot cores as seen in Figure \ref{abu3}. 
The dissimilarity is also seen in the $N$(X)/$N$(CH$_3$OH) comparison between SMM1 and ST16 as shown in Figure \ref{abu2} (b), where the correlation coefficient is calculated to be 0.69. 

\citet{ST20} argue that SO$_2$ will be a good tracer of low-metallicity hot core chemistry, because (i) it is commonly detected in LMC hot cores with similar abundances, and (ii) it is originated from a compact hot core region. 
SO also shows similar abundances within LMC hot cores. 
In WB89-789 SMM1, however, the abundances of SO$_2$ and SO relative to H$_2$ are lower by a factor of 28 and 5 compared with LMC hot cores. 
The measured rotation temperatures of SO$_2$ are similar between those hot cores, i.e., 166 K (SO$_2$) for SMM1, 232 K (SO$_2$) and 86 K ($^{34}$SO$_2$) for ST16, 190 K (SO$_2$) and 95 K ($^{34}$SO$_2$) for ST11. 
The SO$_2$ column densities for ST16 and ST11 are estimated from $^{34}$SO$_2$, while that for SMM1 is from SO$_2$. 
However, the SO$_2$ column density of SMM1 increases only by a factor of up to three when it is estimated from $^{34}$SO$_2$ (see Section \ref{sec_x}). 
Thus the low SO$_2$ abundance in the outer Galactic hot core would not be due to the optical thickness. 

In contrast to the S-O bond bearing species, the C-S bond bearing species such as CS, H$_2$CS , and OCS do not show significant abundance decrease in WB89-789 SMM1. 
Thus it is not straightforward to attribute the low abundance of SO$_2$ (and perhaps SO) to the low elemental abundance ratio of sulfur in the outer Galaxy. 
Hot core chemistry models suggest that SO$_2$ is mainly produced in high-temperature gas-phase reactions in warm gas, using H$_2$S sublimated from ice mantles \citep{Cha97, NM04}. 
This also applies to the SO$_2$ formation in low-metallicity sources as shown in astrochemical simulations for LMC hot cores \citep{ST20}. 
We speculate that the different behavior of SO$_2$ in outer Galaxy and LMC hot cores may be related to differences in the evolutionary stage of hot cores. 
A different luminosity of host protostars may also contribute to the different sulfur chemistry; i.e., $\sim$8 $\times$ 10$^3$ L$_{\sun}$ for WB89-789-SMM1, while several $\times$ 10$^5$ L$_{\sun}$ for LMC hot cores. 
A different cosmic-ray ionization rate between the outer Galaxy and the LMC may also affect the chemical evolution, although the rate is not known for the outer Galaxy.

Among nitrogen-bearing molecules, NO shows interesting behavior in LMC hot cores. 
After corrected for the metallicity, NO is overabundant in LMC hot cores compared with Galactic counterparts despite the low elemental abundance of nitrogen in the LMC \citep{ST20}. 
Only NO shows such behavior among the nitrogen-bearing molecules observed in LMC hot cores. 
In WB89-789 SMM1, however, such an overabundance of NO is not observed. 
The NO abundance of SMM1 is 1.6 $\times$ 10$^{-9}$ for a 0.1 pc region data. 
This is a factor of five lower than a typical NO abundance in Galactic high-mass hot cores \citep[8 $\times$ 10$^{-9}$,][]{Ziu91}, which is consistent with a factor of four lower metallicity in WB89-789. 
The present high-spatial resolution data have revealed that NO does not mainly arise from a hot core region, as shown in Figure \ref{images1}. 
It has an intensity peak at the south part of the hot core, where low-excitation lines of SO and SO$_2$ also have a secondary peak (Section \ref{sec_disc_dist}). 
Thus, shock chemistry or photochemistry, rather than high-temperature chemistry, would contribute to the production of NO in low-metallicity protostellar cores. 
In that case, a lower luminosity of SMM1 than those of LMC hot cores may contribute to the different behavior of NO.

For other nitrogen-bearing molecules, HNCO and CH$_3$CN, a clear difference is not identified between outer Galactic and LMC hot cores, although the number of data points is limited and the abundance uncertainty is large. 
The reason of the unusually low abundance of SiO in SMM1 is unknown. 
It may be related to different shock conditions or grain compositions, because dust sputtering by shock is mainly responsible for the production of SiO gas. 

Formation of COMs is one of the important standpoints for low-metallicity hot core chemistry. 
It is reported that CH$_3$OH show a large abundance variation in LMC hot cores \citep{ST20}. 
There are organic-poor hot cores such as ST11 and ST16, while N113 A1 and B3 are organic-rich. 
The CH$_3$OH abundance of WB89-789 SMM1 is higher than those of any known LMC hot cores. 
The abundances of HCOOCH$_3$ and CH$_3$OCH$_3$ in SMM1 are comparable with those of an organic-rich LMC hot core, N113 A1. 
Th detection of many other COMs in SMM1 suggests the source have experienced rich organic chemistry despite its low-metallicity nature. 

Astrochemical simulations for LMC hot cores suggest that dust temperature at the initial ice-forming stage have a major effect on the abundance of CH$_3$OH gas in the subsequent hot core stage \citep{Ach18,ST20}. 
Simulations of grain surface chemistry dedicated to the LMC environment also suggest that dust temperature is one of the key parameters for the formation of CH$_3$OH in dense cores \citep{Ach15,Pau18}. 
This is because (i) CH$_3$OH is mainly formed by the grain surface reaction, and (ii) the hydrogenation reaction of CO, which is a dominant pathway for the CH$_3$OH formation, is sensitive to the dust temperature due to the high volatility of atomic hydrogen. 
For this reason, it is inferred that organic-rich hot cores had experienced a cold stage ($\lesssim$ 10K) that is sufficient for the CH$_3$OH formation before the hot core stage, while organic-poor ones might have missed such a condition for some reason. 
Alternatively, the slight difference in the hot core's evolutionary stage may contribute to the CH$_3$OH abundance variation, because the high-temperature gas-phase chemistry is rapid and it can decrease CH$_3$OH gas at a late stage \citep[e.g.,][]{NM04,Gar06,Vas13,Bal15}. 

Low-metallicity hot core chemistry simulations in \citet{ST20} argue that the maximum achievable abundances of CH$_3$OH gas in a hot core stage significantly decrease as the visual extinction of the initial ice-forming stage decreases. 
On the other hand, the simulations show that the CH$_3$OH gas abundance is simply metallicity-scaled if the initial ice-forming stage is sufficiently shielded. 
In a well-shielded initial condition, the grain surface is cold enough to trigger the CO hydrogenation, and the resultant CH$_3$OH abundance is roughly regulated by the elemental abundances. 
The observed metallicity-scaled chemistry of COMs in WB89-789 SMM1 implies that the source had experienced such an initial condition before the hot core stage. 

Deuterium chemistry is widely used in interpreting chemical and physical history of interstellar molecules \citep[e.g.,][]{Cas12,Cec14}. 
The measured CH$_2$DOH/CH$_3$OH ratio in WB89-789 SMM1 is 1.1 $\pm$ 0.2 $\%$, which is comparable to the higher end of those ratios observed in high-mass protostars and the lower end of those in low-mass protostars \citep[e.g., see Fig.2 in][]{Dro21}. 
The ratio is orders of magnitude higher than the deuterium-to-hydrogen ratio in the solar neighborhood \citep[2 $\times$ 10$^{-5}$;][]{Lin06,Pro10} and that in the big-bang nucleosynthesis \citep[3 $\times$ 10$^{-5}$;][references therein]{Bur02}. 
This suggests that the efficient deuterium fractionation occurred upon the formation of CH$_3$OH in SMM1. 
The D$_2$CO/HDCO ratio is 45 $\pm$ 10 $\%$, which is comparable to those observed in low-mass and high-mass protostars \citep[e.g.,][]{Zah21}. 
This would suggest that physical conditions for deuterium fractionation could be similar between WB89-789 SMM1 and inner Galactic protostars. 
Note that higher spatial resolution observations and detailed multiline analyses would affect the measured abundance of deuterated species as \replaced{shown}{reported} in \citet{Per18} for the case of a nearby low-mass protostar. 
The H$_2$CO column density derived in this work may be a lower limit because the line is often optically thick, thus we do not discuss the abundance ratio relative to H$_2$CO. 

It is known that the deuterium fractionation efficiently proceeds at low temperature \citep[e.g.,][]{Rob03,Cas12,Taq14,Fur16}. 
This is because the key reaction for the trigger of deuterium fractionation, H$^+_3$ + HD $\rightarrow$ H$_2$D$^+$ + H$_2$ + 232 K, is exothermic and its backward reaction cannot \added{efficiently} proceed below 20 K. 
In addition, gaseous neutral species such as CO and O efficiently destruct H$_2$D$^+$, thus their depletion at low temperature further enhances the deuterium fractionation \citep[e.g.,][]{Cas12}. 
A sign of high deuterium fractionation observed in WB89-789 SMM1 suggests that the source had experienced such a cold environment during its formation. 
This picture is consistent with the implication obtained from the metallicity-scaled chemistry of COMs, which also suggests the occurrence of a cold and well-shielded initial condition as discussed above. 

Although the low metallicity is common between the outer Galaxy and the LMC, their star-forming environments would be different; the LMC has more harsh environments as inferred from active massive star formation over the whole galaxy, while that for the outer Galaxy might be quiescent due to its low star formation activity. 
Those environmental differences need to be taken into account for further understanding of the chemical evolution of star-forming regions at low metallicity. 
Future extensive survey of protostellar objects towards the outer Galaxy is thus vitally important for further discussion. 
Astrochemical simulations dedicated to the environment of the outer Galaxy, and the application to lower-mass protostars, are also important.

\subsection{Another embedded protostar traced by high-velocity SiO gas} \label{sec_disc_SiO} 
\begin{figure}[tpb!]
\begin{center}
\includegraphics[width=8.5cm]{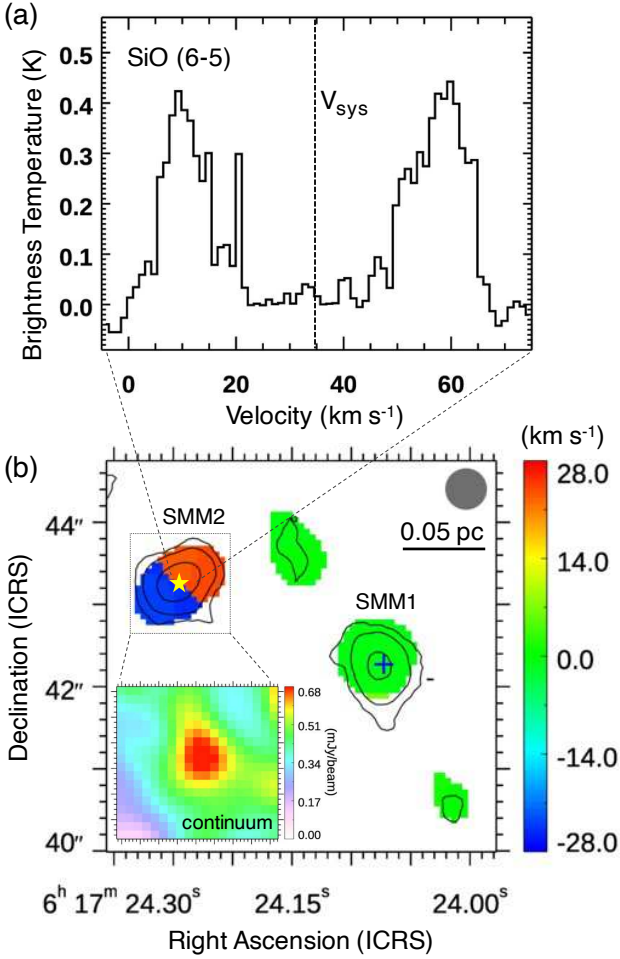}
\caption{
(a) SiO(6-5) spectrum of WB89-789-SMM2. 
The dotted line indicates a systemic velocity of 34.5 km s$^{-1}$. 
High-velocity ($V_{sys}$ $\pm$25 km s$^{-1}$) SiO components are seen at the blue-/red-shifted sides of the systemic velocity. 
(b) Velocity map (moment 1) of the SiO(6-5) line. 
The color scale indicates the offset velocity relative to the systemic velocity. 
\added{Low signal-to-noise ratio regions (S/N $<$5) are masked. }
Gray contours represent the intensity distribution of SiO(6-5) integrated from 0 to 60 km s$^{-1}$, and the contour levels are 1.5$\sigma$, 4$\sigma$, and 12$\sigma$ of the rms level. 
The yellow star indicates the SiO center of SMM2, while the blue cross indicates the hot core position (SMM1). 
The subset panel shows the 1200 $\mu$m continuum image for a 1$\farcs$2 $\times$ 1$\farcs$2 region centered at SMM2.  
See Section \ref{sec_disc_SiO} for details. 
}
\label{moment1_SiO}
\end{center}
\end{figure}

We have also detected a compact source associated with high-velocity SiO gas at the east side of WB89-789 SMM1. 
Hereafter, we refer to this source as WB89-789-SMM2. 
According to the SiO emission, the source is located at RA = 06$^\mathrm{h}$17$^\mathrm{m}$24$\fs$246 and Dec = 14$^\circ$54$\arcmin$43$\farcs$25 (ICRS), which is 2$\farcs$7 (0.14 pc) away from SMM2. 
Figure \ref{moment1_SiO}(a) shows the SiO(6-5) spectrum extracted from a 0$\farcs$6 diameter region centered at the above position. 
The SiO line is largely shifted to the blue and red sides relative to the systemic velocity in a symmetric fashion. 
The peaks of the shifted emission are located at $V_{sys}$ $\pm$ 25 km s$^{-1}$. 

Figure \ref{moment1_SiO}(b) shows a velocity map and integrated intensity distribution of SiO(6-5). 
In the figure, to focus on SiO in WB89-789-SMM2, the intensity is integrated over much wider velocity range (0--60 km s$^{-1}$) compared with that adopted in Figure \ref{images1} (31--38 km s$^{-1}$). 
The velocity map clearly indicates that the velocity structure of SiO in SMM2 is spatially symmetric to the SiO center. 
At this position, a local peak is seen in 1200 $\mu$m continuum as shown in the figure, suggesting the presence of an embedded source. 
SMM2 does not show any emission lines of COMs, and no alternative molecular lines are identified at the \replaced{positions}{frequencies} of doppler-shifted SiO emission. 
Also taking into account the clear spectral and spatial symmetry, the observed lines must be attributed to high-velocity SiO gas. 

The spectral characteristics of the observed high-velocity SiO resemble those of extremely high velocity (EHV) outflows observed in Class 0 protostars \citep{Bac91,Taf10,Taf15,Tyc19}. 
The EHV flows are known to appear as a discrete high-velocity ($V$ $\gtrsim$30 km s$^{-1}$) peak, and observed in the youngest stage of star formation \citep[][references therein]{Bac96,Mat19}. 
The EHV flows extends up to several thousands au from the central protostar in SiO, and usually have collimated bipolar structures \citep[e.g.,][]{Bac91,Hir10,Tyc19,Mat19}. 
The beam size of the present data is about 5000 au, thus such structures will not be \added{fully} spatially resolved. 
Actually, a symmetric spatial distribution of blue-/red-shifted SiO is only marginally resolved into two beam size regions (Fig. \ref{moment1_SiO}(b)). 
A spatial extent of SiO emission is about 1$\arcsec$ (0.052 pc). 
Assuming an outflow velocity of 25 km s$^{-1}$, we estimate a dynamical timescale of EHV flows to be at least 2000 years. 
This is roughly consistent with dynamical timescales of other EHV sources, which range from a few hundred to a few thousand years \citep[][references therein]{Bac96}. 

A 1200 $\mu$m continuum flux in a 0$\farcs$6 diameter region centered at SMM2 is 0.60 $\pm$ 0.05 mJy/beam. 
Assuming $T_{d}$ = 20 K, we obtain $N_{\mathrm{H_2}}$ = 3.2 $\times$ 10$^{23}$ cm$^{-2}$. 
This is equivalent to a gas number density of $n_{\mathrm{H_2}}$ = 4.9 $\times$ 10$^6$ cm$^{-3}$. 
If we assume a higher $T_{d}$, i.e. 40 K, then the derived column density is 2.5 times lower than the \replaced{above estimate}{20 K case}. 
In either case, the continuum data suggests the presence of high-density gas at this position. 

\added{
A column density and fractional abundance of SiO gas at the above position is estimated to be $N(\mathrm{SiO)}$ $\sim$ 2 $\times$ 10$^{13}$ cm$^{-2}$ and $N(\mathrm{SiO)}$/$N_{\mathrm{H_2}}$ $\sim$ 6 $\times$ 10$^{-11}$, where we assume optically thin emission in the LTE and the gas/dust temperature of 20 K. 
The fractional abundance will be two times higher if we assume the gas/dust temperature of 10 K or 40 K. 
The SiO abundance in SMM2 is at least 30 times higher than that observed in SMM1. 
The observed enhancement of SiO in SMM2 would be related to shock chemistry triggered by EHV outflows. 
}

Previous single-dish observations of CO detected extended ($\sim$20$\arcsec$) molecular outflows in the WB89-789 region \citep{Bra94,Bra07}. 
The center of the outflow gas coincides with the position of the IRAS source (IRAS 06145+1455; 06$^\mathrm{h}$17$^\mathrm{m}$24$\fs$2, 14$\arcdeg$54$\arcmin$42$\arcsec$, J2000). 
This position is consistent with those of SMM1 or SMM2, given the large beam size of CO(3-2) observations (14$\arcsec$) in \citet{Bra07}. 
The observed CO outflow gas has an extended blue-shifted component (20 $<$ $V_{LSR}$ $<$ 31 km $s^{-1}$) towards the south-east direction from the center, while a red-shifted component (37 $<$ $V_{LSR}$ $<$ 55 km $s^{-1}$) is extended towards the north-west direction \citep[see Figure 9 in ][]{Bra07}. 
This outflow direction coincides with that of the high-velocity SiO outflows observed in this work. 
The SiO outflows from SMM2 may have a common origin with the large-scale CO outflows. 

In summary, it is likely that a compact, high-density, and embedded object is located at the position of WB89-789-SMM2. 
Presumably, a protostar associated with SMM2 is driving the observed high-velocity SiO gas flows. 
Its short dynamical timescale and similarity with EHV flows suggest that the object is at the youngest stage of star formation (Class 0/I). 
Non-detection of warm gas emission also supports its young nature. 
We note that the detailed structure of high-velocity SiO gas is not \added{fully} spatially resolved, and CO lines, which often trace high-velocity outflows, are not covered in the present data. 
Future high-spatial resolution observations of CO and other outflow tracers are key to further clarify the nature of WB89-789-SMM2.

\section{Summary} \label{sec_sum} 
The extreme outer Galaxy is an excellent laboratory to study star formation and interstellar medium in \replaced{a low-metallicity and primitive Galactic environment. }{a Galactic low-metallicity environment. }
The following conclusions are obtained in this work. 

\begin{enumerate}
\item 
A hot molecular core is for the first time detected in the extreme outer Galaxy (WB89-789-SMM1), based on submillimeter observations with ALMA towards the WB89-789 star-forming region located at the galactocentric distance of 19 kpc. 

\item 
A variety of carbon-, oxygen-, nitrogen-, sulfur-, and silicon-bearing species, including complex organic molecules containing up to nine atoms and larger than CH$_3$OH, are detected towards a warm ($>$100 K) and compact ($<$ 0.03 pc) region associated with a protostar ($\sim$8 $\times$ 10$^3$ L$_{\sun}$). 
The results suggest that a great molecular complexity exists even in a \replaced{primitive}{lower metallicity} environment of the extreme outer Galaxy. 

\item 
\added{For deuterated species, we have detected HDO, HDCO, D$_2$CO, and CH$_2$DOH. 
HDO and CH$_2$DOH arise from a compact and high-temperature ($T$$_{rot}$ = 155--220 ) region, while HDCO and D$_2$CO are in lower temperature ($T$$_{rot}$ $\sim$ 40 K) and slightly extended. 
The measured ratios of CH$_2$DOH/CH$_3$OH and D$_2$CO/HDCO are 1.1 $\pm$ 0.2 $\%$ and 45 $\pm$ 10 $\%$, respectively. 
}

\item 
Fractional abundances of CH$_3$OH and other COMs relative to H$_2$ generally scale with the metallicity of WB89-789, which is a factor of four lower than the solar value. 

\item 
A comparison of fractional abundances of COMs relative to the CH$_3$OH column density between the outer Galactic hot core and a Galactic intermediate-mass hot core show a remarkable similarity. 
The results suggest the metallicity-scaled chemistry for the formation of COMs in this source. 
CH$_3$OH is an important parental molecule for the COMs formation even in a \replaced{low-metallicity}{lower metallicity} environment. 

\item 
On the other hand, the molecular abundances of the present hot core do not resemble those of LMC hot cores. 
We speculate that different luminosities or star-forming environments between outer Galactic and LMC hot cores may contribute to this. 

\item 
According to astrochemical simulations of low-metallicity hot cores, the observed metallicity-scaled chemistry of COMs in WB89-789-SMM1 implies that the source had experienced well-shielded and cold ice-forming stage before the hot core stage. 

\item 
We have also detected another compact source (WB89-789-SMM2) associated with high-velocity SiO gas ($V_{sys}$ $\pm$ 25 km s$^{-1}$) in the same region. 
The characteristics of the source resemble those of EHV outflows observed in Class 0 protostars. 
Physical properties and dynamical timescale of this outflow source are discussed. 
\end{enumerate}

This paper makes use of the following ALMA data: ADS/JAO.ALMA$\#$2017.1.01002.S \added{and 2018.1.00627.S. }
ALMA is a partnership of ESO (representing its member states), NSF (USA) and NINS (Japan), together with NRC (Canada), MOST and ASIAA (Taiwan), and KASI (Republic of Korea), in cooperation with the Republic of Chile. 
The Joint ALMA Observatory is operated by ESO, AUI/NRAO and NAOJ. 
This work has made extensive use of the Cologne Database for Molecular Spectroscopy and the molecular database of the Jet Propulsion Laboratory. 
This work makes use of data products from the Two Micron All Sky Survey, which is a joint project of the University of Massachusetts and the Infrared Processing and Analysis Center/California Institute of Technology, funded by the National Aeronautics and Space Administration and the National Science Foundation.
This work was supported by JSPS KAKENHI Grant Number \replaced{19H05067 and 21H00037}{19H05067, 21H00037, and 21H01145}. 
\added{Finally, we would like to thank an anonymous referee for insightful comments, which substantially improved this paper.  }

%




\software{CASA \citep{McM07})}




%

\clearpage

\appendix

\restartappendixnumbering

\section{Measured line parameters}
Tables \ref{tab_lines1}--\ref{tab_lines9} summarize measured line parameters (see Section \ref{sec_spc} for details). 
The tabulated uncertainties and upper limits are of 2$\sigma$ level and do not include systematic errors due to continuum subtraction. 
Upper limits are estimated assuming $\Delta$$V$ = 4 km s$^{-1}$.

\startlongtable
\begin{deluxetable*}{ l c c c c c c c c c }
\tablecaption{Line Parameters for HDO, H$^{13}$CO$^+$, HC$^{18}$O$^+$, CCH, c-C$_3$H$_2$, H$_2$CO, HDCO, D$_2$CO \label{tab_lines1}}
\tablewidth{0pt}
\tabletypesize{\scriptsize} 
\tablehead{
\colhead{Molecule}                       & \colhead{Transition}                                                             &       \colhead{$E_{u}$} &       \colhead{Frequency} &        \colhead{$T_{br}$} &     \colhead{$\Delta$$V$} &     \colhead{$\int T_{br} dV$} &       \colhead{$V_{LSR}$} &        \colhead{RMS} &       \colhead{Note} \\
\colhead{ }                              & \colhead{ }                                                                      &        \colhead{(K)} &           \colhead{(GHz)} &             \colhead{(K)} &          \colhead{(km/s)} &             \colhead{(K km/s)} &          \colhead{(km/s)} &        \colhead{(K)} &           \colhead{}
}
\startdata
 HDO                                     &  2$_{1,1}$--2$_{1,2}$                                                                                                    &   95       & 241.56155       &   0.99 $\pm$   0.03  &    4.8          & 5.05 $\pm$ 0.29      &  34.6           & 0.04 &    \nodata \\
 HDO                                     &  7$_{3,4}$--6$_{4,3}$                                                                                                    &  837       & 241.97357       &   0.28 $\pm$   0.02  &    2.1          & 0.63 $\pm$ 0.12      &  34.1           & 0.04 &    \nodata \\
 H$^{13}$CO$^+$                          &  3--2                                                                                                                    &   25       & 260.25534       &   5.85 $\pm$   0.02  &    2.3          & 14.23 $\pm$ 0.13     &  34.1           & 0.03 &    \nodata \\
 HC$^{18}$O$^+$                          &  4--3                                                                                                                    &   41       & 340.63069       &   0.64 $\pm$   0.05  &    1.9          & 1.29 $\pm$ 0.20      &  34.1           & 0.07 &    \nodata \\
 CCH                                     &  N = 4--3, J = $\frac{9}{2}$--$\frac{7}{2}$, F = 5--4                                                                    &   42       & 349.33771       &   4.39 $\pm$   0.04  &    2.6          & 12.32 $\pm$ 0.24     &  33.7           & 0.08 &    (1) \\
 CCH                                     &  N = 4--3, J = $\frac{7}{2}$--$\frac{5}{2}$, F = 4--3                                                                    &   42       & 349.39928       &   3.54 $\pm$   0.04  &    2.8          & 10.49 $\pm$ 0.26     &  33.7           & 0.08 &    (1) \\
\enddata
\tablecomments{
(1) Blend of two hyperfine components. 
\added{(2) Tentative detection. }
}
\end{deluxetable*}

\startlongtable
\begin{deluxetable*}{ l c c c c c c c c c }
\tablecaption{Line Parameters for N-bearing molecules \label{tab_lines2}}
\tablewidth{0pt}
\tabletypesize{\scriptsize} 
\tablehead{
\colhead{Molecule}                       & \colhead{Transition}                                                             &       \colhead{$E_{u}$} &       \colhead{Frequency} &        \colhead{$T_{br}$} &     \colhead{$\Delta$$V$} &     \colhead{$\int T_{br} dV$} &       \colhead{$V_{LSR}$} &        \colhead{RMS} &       \colhead{Note} \\
\colhead{ }                              & \colhead{ }                                                                      &        \colhead{(K)} &           \colhead{(GHz)} &             \colhead{(K)} &          \colhead{(km/s)} &             \colhead{(K km/s)} &          \colhead{(km/s)} &        \colhead{(K)} &           \colhead{}
}
\startdata
 H$^{13}$CN                              &  3--2                                                                                                                    &   25       & 259.01180       &   3.25 $\pm$   0.02  &    4.2          & 14.45 $\pm$ 0.19     &  34.3           & 0.03 &    \nodata \\
 HC$^{15}$N                              &  3--2                                                                                                                    &   25       & 258.15700       &   1.95 $\pm$   0.02  &    3.6          & 7.45 $\pm$ 0.19      &  34.3           & 0.03 &    \nodata \\
 CN                                      &  N = 3--2, J = $\frac{5}{2}$--$\frac{5}{2}$, F = $\frac{3}{2}$--$\frac{3}{2}$                                            &   33       & 339.44678       &   0.30 $\pm$   0.04  &    2.6          & 0.82 $\pm$ 0.22      &  34.6           & 0.07 &    \nodata \\
 CN                                      &  N = 3--2, J = $\frac{5}{2}$--$\frac{5}{2}$, F = $\frac{5}{2}$--$\frac{5}{2}$                                            &   33       & 339.47590       &   0.39 $\pm$   0.04  &    1.7          & 0.71 $\pm$ 0.17      &  34.0           & 0.07 &    \nodata \\
 CN                                      &  N = 3--2, J = $\frac{5}{2}$--$\frac{5}{2}$, F = $\frac{7}{2}$--$\frac{7}{2}$                                            &   33       & 339.51664       &   0.73 $\pm$   0.04  &    1.8          & 1.38 $\pm$ 0.18      &  34.1           & 0.07 &    \nodata \\
 CN                                      &  N = 3--2, J = $\frac{5}{2}$--$\frac{3}{2}$, F = $\frac{5}{2}$--$\frac{5}{2}$                                            &   33       & 340.00813       &   0.95 $\pm$   0.04  &    2.1          & 2.16 $\pm$ 0.20      &  33.9           & 0.07 &    \nodata \\
\enddata
\tablecomments{
(1) Blend of three hyperfine components. 
(2) Blend of three two components. 
(3) Blend with CH$_3$OH. 
(4) Blend with HCOOCH$_3$. 
(5) Blend of four hyperfine components. 
}
\end{deluxetable*}

\startlongtable
\begin{deluxetable*}{ l c c c c c c c c c }
\tablecaption{Line Parameters for Si- and S-bearing molecules \label{tab_lines3}}
\tablewidth{0pt}
\tabletypesize{\scriptsize} 
\tablehead{
\colhead{Molecule}                       & \colhead{Transition}                                                             &       \colhead{$E_{u}$} &       \colhead{Frequency} &        \colhead{$T_{br}$} &     \colhead{$\Delta$$V$} &     \colhead{$\int T_{br} dV$} &       \colhead{$V_{LSR}$} &        \colhead{RMS} &       \colhead{Note} \\
\colhead{ }                              & \colhead{ }                                                                      &        \colhead{(K)} &           \colhead{(GHz)} &             \colhead{(K)} &          \colhead{(km/s)} &             \colhead{(K km/s)} &          \colhead{(km/s)} &        \colhead{(K)} &           \colhead{}
}
\startdata
 SiO                                     &  6--5                                                                                                                    &   44       & 260.51801       &   0.64 $\pm$   0.02  &    2.7          & 1.86 $\pm$ 0.13      &  34.1           & 0.03 &    \nodata \\
 SO                                      &  $N_J$ = 6$_{6}$--5$_{5}$                                                                                                &   56       & 258.25583       &   5.34 $\pm$   0.02  &    3.6          & 20.20 $\pm$ 0.19     &  34.1           & 0.03 &    \nodata \\
 SO                                      &  $N_J$ = 3$_{3}$--2$_{3}$                                                                                                &   26       & 339.34146       &   0.49 $\pm$   0.04  &    3.4          & 1.76 $\pm$ 0.34      &  34.0           & 0.07 &    \nodata \\
 SO                                      &  $N_J$ = 8$_{7}$--7$_{6}$                                                                                                &   81       & 340.71416       &   3.85 $\pm$   0.04  &    3.9          & 15.81 $\pm$ 0.39     &  34.2           & 0.07 &    \nodata \\
 $^{34}$SO                               &  $N_J$ = 3$_{3}$--2$_{3}$                                                                                                &   25       & 337.89225       &             $<$0.15  &    \nodata      &             $<$0.6   &    \nodata      & 0.07 &    \nodata \\
 $^{34}$SO                               &  $N_J$ = 8$_{9}$--7$_{8}$                                                                                                &   77       & 339.85727       &   0.69 $\pm$   0.04  &    4.4          & 3.26 $\pm$ 0.38      &  33.0           & 0.07 &    \nodata \\
\enddata
\tablecomments{
(1) Blend of four hyperfine components. 
(2) The integrated intensity is calculated by directly integrating the spectrum. 
(3) Partial blend with HNCO. 
(4) Blend with HCOOCH$_3$. 
(5) Blend with CH$_3$OH. 
}
\end{deluxetable*}

\startlongtable
\begin{deluxetable*}{ l c c c c c c c c c }
\tablecaption{Line Parameters for CH$_3$OH, $^{13}$CH$_3$OH, and CH$_2$DOH \label{tab_lines4}}
\tablewidth{0pt}
\tabletypesize{\scriptsize} 
\tablehead{
\colhead{Molecule}                       & \colhead{Transition}                                                             &       \colhead{$E_{u}$} &       \colhead{Frequency} &        \colhead{$T_{br}$} &     \colhead{$\Delta$$V$} &     \colhead{$\int T_{br} dV$} &       \colhead{$V_{LSR}$} &        \colhead{RMS} &       \colhead{Note} \\
\colhead{ }                              & \colhead{ }                                                                      &        \colhead{(K)} &           \colhead{(GHz)} &             \colhead{(K)} &          \colhead{(km/s)} &             \colhead{(K km/s)} &          \colhead{(km/s)} &        \colhead{(K)} &           \colhead{}
}
\startdata
 CH$_3$OH                                &  25$_{3}$ A$^-$--25$_{2}$ A$^+$                                                                                          &  804       & 241.58876       &   0.72 $\pm$   0.02  &    4.5          & 3.49 $\pm$ 0.25      &  34.5           & 0.04 &    \nodata \\
 CH$_3$OH                                &  5$_{0}$ E--4$_{0}$ E                                                                                                    &   48       & 241.70016       &   2.45 $\pm$   0.03  &    4.0          & 10.57 $\pm$ 0.24     &  34.2           & 0.04 &    \nodata \\
 CH$_3$OH                                &  5$_{-1}$ E--4$_{-1}$ E                                                                                                  &   40       & 241.76723       &   3.70 $\pm$   0.03  &    3.2          & 12.53 $\pm$ 0.20     &  34.2           & 0.04 &    \nodata \\
 CH$_3$OH                                &  5$_{0}$ A$^+$--4$_{0}$ A$^+$                                                                                            &   35       & 241.79135       &   4.15 $\pm$   0.03  &    3.0          & 13.29 $\pm$ 0.19     &  34.1           & 0.04 &    \nodata \\
 CH$_3$OH                                &  5$_{4}$ A$^-$--4$_{4}$ A$^-$                                                                                            &  115       & 241.80652       &   1.46 $\pm$   0.02  &    4.3          & 6.65 $\pm$ 0.23      &  34.7           & 0.04 &    (1) \\
 CH$_3$OH                                &  5$_{-4}$ E--4$_{-4}$ E                                                                                                  &  123       & 241.81325       &   1.33 $\pm$   0.02  &    4.3          & 6.13 $\pm$ 0.23      &  34.7           & 0.04 &    \nodata \\
\enddata
\tablecomments{
(1) Blend of two CH$_3$OH lines with similar spectroscopic constants. 
(2) Blend of three CH$_3$OH lines with similar spectroscopic constants. 
(3) Possible blend with C$_2$H$_5$OH. 
(4) Blend with CH$_3$CN. 
(5) Blend with HCOOCH$_3$. 
(6) Blend with $^{34}$SO. 
(7) Partial blend with HC$^{15}$N. 
(8) Possible blend with HDCO. 
(9) Blend with CH$_3$OCH$_3$. 
(10) Partial blend with CN. 
\added{(11) Tentative detection. }
(12) Blend of two CH$_2$DOH lines with similar spectroscopic constants. 
}
\end{deluxetable*}

\startlongtable
\begin{deluxetable*}{ l c c c c c c c c c }
\tablecaption{Line Parameters for C$_2$H$_5$OH \label{tab_lines5}}
\tablewidth{0pt}
\tabletypesize{\scriptsize} 
\tablehead{
\colhead{Molecule}                       & \colhead{Transition}                                                             &       \colhead{$E_{u}$} &       \colhead{Frequency} &        \colhead{$T_{br}$} &     \colhead{$\Delta$$V$} &     \colhead{$\int T_{br} dV$} &       \colhead{$V_{LSR}$} &        \colhead{RMS} &       \colhead{Note} \\
\colhead{ }                              & \colhead{ }                                                                      &        \colhead{(K)} &           \colhead{(GHz)} &             \colhead{(K)} &          \colhead{(km/s)} &             \colhead{(K km/s)} &          \colhead{(km/s)} &        \colhead{(K)} &           \colhead{}
}
\startdata
 C$_2$H$_5$OH                            &  14$_{11,3}$--13$_{11,2}$                                                                                                &  297       & 242.17548       &   0.15 $\pm$   0.02  &    5.4          & 0.87 $\pm$ 0.40      &  34.7           & 0.04 &    (1) \\
 C$_2$H$_5$OH                            &  14$_{9,5}$--13$_{9,4}$                                                                                                  &  248       & 242.22129       &   0.16 $\pm$   0.02  &    5.5          & 0.93 $\pm$ 0.29      &  33.7           & 0.04 &    (1) \\
 C$_2$H$_5$OH                            &  14$_{8,6}$--13$_{8,5}$                                                                                                  &  228       & 242.27115       &   0.13 $\pm$   0.02  &    4.1          & 0.57 $\pm$ 0.23      &  34.6           & 0.04 &    (1) \\
 C$_2$H$_5$OH                            &  14$_{7,8}$--13$_{7,7}$                                                                                                  &  209       & 242.34984       &   0.30 $\pm$   0.02  &    4.1          & 1.30 $\pm$ 0.23      &  35.7           & 0.04 &    (1) \\
 C$_2$H$_5$OH                            &  14$_{10,4}$--13$_{10,3}$                                                                                                &  266       & 242.42987       &   0.36 $\pm$   0.02  &    1.6          & 0.62 $\pm$ 0.08      &  35.3           & 0.04 &    (2) \\
 C$_2$H$_5$OH                            &  14$_{6,9}$--13$_{6,8}$                                                                                                  &  193       & 242.47550       &   0.35 $\pm$   0.02  &    4.3          & 1.60 $\pm$ 0.23      &  34.6           & 0.04 &    (1) \\
\enddata
\tablecomments{
(1) Blend of two C$_2$H$_5$OH lines with similar spectroscopic constants. 
(2) Blend of four C$_2$H$_5$OH lines with similar spectroscopic constants. 
\added{(3) Tentative detection. }
(4) Blend of three C$_2$H$_5$OH lines with similar spectroscopic constants. 
(5) Possible blend with CH$_3$CHO. 
}
\end{deluxetable*}

\startlongtable
\begin{deluxetable*}{ l c c c c c c c c c }
\tablecaption{Line Parameters for HCOOCH$_3$ \label{tab_lines6}}
\tablewidth{0pt}
\tabletypesize{\scriptsize} 
\tablehead{
\colhead{Molecule}                       & \colhead{Transition}                                                             &       \colhead{$E_{u}$} &       \colhead{Frequency} &        \colhead{$T_{br}$} &     \colhead{$\Delta$$V$} &     \colhead{$\int T_{br} dV$} &       \colhead{$V_{LSR}$} &        \colhead{RMS} &       \colhead{Note} \\
\colhead{ }                              & \colhead{ }                                                                      &        \colhead{(K)} &           \colhead{(GHz)} &             \colhead{(K)} &          \colhead{(km/s)} &             \colhead{(K km/s)} &          \colhead{(km/s)} &        \colhead{(K)} &           \colhead{}
}
\startdata
 HCOOCH$_3$                              &  20$_{4,17}$--19$_{4,16}$ A $\nu_t$ = 1                                                                                  &  322       & 242.61007       &   0.35 $\pm$   0.02  &    2.4          & 0.92 $\pm$ 0.14      &  34.3           & 0.04 &    \nodata \\
 HCOOCH$_3$                              &  19$_{5,14}$--18$_{5,13}$ A                                                                                              &  130       & 242.89603       &   0.74 $\pm$   0.02  &    3.6          & 2.85 $\pm$ 0.19      &  34.5           & 0.04 &    \nodata \\
 HCOOCH$_3$                              &  19$_{4,15}$--18$_{4,14}$ A $\nu_t$ = 1                                                                                  &  313       & 244.06667       &   0.35 $\pm$   0.02  &    3.4          & 1.28 $\pm$ 0.18      &  34.4           & 0.04 &    \nodata \\
 HCOOCH$_3$                              &  20$_{10,10}$--19$_{10,9}$ E $\nu_t$ = 1                                                                                 &  378       & 244.11242       &   0.13 $\pm$   0.02  &    4.0          & 0.56 $\pm$ 0.22      &  33.9           & 0.04 &    \nodata \\
 HCOOCH$_3$                              &  20$_{12,8}$--19$_{12,7}$ A $\nu_t$ = 1                                                                                  &  407       & 244.19830       &   0.17 $\pm$   0.02  &    4.8          & 0.89 $\pm$ 0.25      &  34.5           & 0.04 &    (1) \\
 HCOOCH$_3$                              &  20$_{10,11}$--19$_{10,10}$ A $\nu_t$ = 1                                                                                &  377       & 244.52854       &   0.38 $\pm$   0.02  &    3.0          & 1.21 $\pm$ 0.16      &  34.7           & 0.04 &    (1) \\
\enddata
\tablecomments{
(1) Blend of two HCOOCH$_3$ lines with similar spectroscopic constants. 
(2) Blend of three HCOOCH$_3$ lines with similar spectroscopic constants. 
\added{(3) Tentative detection. }
}
\end{deluxetable*}

\startlongtable
\begin{deluxetable*}{ l c c c c c c c c c }
\tablecaption{Line Parameters for CH$_3$OCH$_3$ \label{tab_lines7}}
\tablewidth{0pt}
\tabletypesize{\scriptsize} 
\tablehead{
\colhead{Molecule}                       & \colhead{Transition}                                                             &       \colhead{$E_{u}$} &       \colhead{Frequency} &        \colhead{$T_{br}$} &     \colhead{$\Delta$$V$} &     \colhead{$\int T_{br} dV$} &       \colhead{$V_{LSR}$} &        \colhead{RMS} &       \colhead{Note} \\
\colhead{ }                              & \colhead{ }                                                                      &        \colhead{(K)} &           \colhead{(GHz)} &             \colhead{(K)} &          \colhead{(km/s)} &             \colhead{(K km/s)} &          \colhead{(km/s)} &        \colhead{(K)} &           \colhead{}
}
\startdata
 CH$_3$OCH$_3$                           &  5$_{3,2}$--4$_{2,3}$ EE                                                                                                 &   26       & 241.52872       &   0.59 $\pm$   0.02  &    5.0          & 3.12 $\pm$ 0.26      &  33.3           & 0.04 &    (1) \\
 CH$_3$OCH$_3$                           &  5$_{3,2}$--4$_{2,3}$ AA                                                                                                 &   26       & 241.53103       &   0.58 $\pm$   0.02  &    5.2          & 3.23 $\pm$ 0.32      &  36.3           & 0.04 &    \nodata \\
 CH$_3$OCH$_3$                           &  21$_{3,18}$--20$_{4,17}$ EE                                                                                             &  226       & 241.63730       &   0.25 $\pm$   0.02  &    4.7          & 1.25 $\pm$ 0.29      &  34.8           & 0.04 &    (1) \\
 CH$_3$OCH$_3$                           &  13$_{1,13}$--12$_{0,12}$ EE                                                                                             &   81       & 241.94654       &   0.92 $\pm$   0.02  &    4.2          & 4.12 $\pm$ 0.23      &  34.5           & 0.04 &    (2) \\
 CH$_3$OCH$_3$                           &  23$_{2,22}$--23$_{1,23}$ EE                                                                                             &  253       & 244.50830       &   0.16 $\pm$   0.02  &    4.2          & 0.72 $\pm$ 0.24      &  33.5           & 0.04 &    \nodata \\
 CH$_3$OCH$_3$                           &  23$_{2,22}$--23$_{1,23}$ AA                                                                                             &  253       & 244.51274       &   0.12 $\pm$   0.02  &    3.3          & 0.40 $\pm$ 0.17      &  34.5           & 0.04 &    \nodata \\
\enddata
\tablecomments{
(1) Blend of three CH$_3$OCH$_3$ lines with similar spectroscopic constants. 
(2) Blend of four CH$_3$OCH$_3$ lines with similar spectroscopic constants. 
\added{(3) Tentative detection. }
}
\end{deluxetable*}

\startlongtable
\begin{deluxetable*}{ l c c c c c c c c c }
\tablecaption{Line Parameters for C$_2$H$_5$CN and NH$_2$CHO \label{tab_lines8}}
\tablewidth{0pt}
\tabletypesize{\scriptsize} 
\tablehead{
\colhead{Molecule}                       & \colhead{Transition}                                                             &       \colhead{$E_{u}$} &       \colhead{Frequency} &        \colhead{$T_{br}$} &     \colhead{$\Delta$$V$} &     \colhead{$\int T_{br} dV$} &       \colhead{$V_{LSR}$} &        \colhead{RMS} &       \colhead{Note} \\
\colhead{ }                              & \colhead{ }                                                                      &        \colhead{(K)} &           \colhead{(GHz)} &             \colhead{(K)} &          \colhead{(km/s)} &             \colhead{(K km/s)} &          \colhead{(km/s)} &        \colhead{(K)} &           \colhead{}
}
\startdata
 C$_2$H$_5$CN                            &  27$_{3,25}$--26$_{3,24}$                                                                                                &  173       & 241.62587       &   0.13 $\pm$   0.02  &    4.9          & 0.69 $\pm$ 0.29      &  33.3           & 0.04 &    \nodata \\
 C$_2$H$_5$CN                            &  27$_{9,18}$--26$_{9,17}$                                                                                                &  253       & 241.93218       &   0.23 $\pm$   0.03  &    3.6          & 0.86 $\pm$ 0.21      &  33.9           & 0.04 &    (1) \\
 C$_2$H$_5$CN                            &  27$_{12,15}$--26$_{12,14}$                                                                                              &  322       & 241.95905       &   0.12 $\pm$   0.02  &    3.1          & 0.40 $\pm$ 0.16      &  34.2           & 0.04 &    (2) \\
 C$_2$H$_5$CN                            &  27$_{8,20}$--26$_{8,19}$                                                                                                &  234       & 241.97045       &   0.30 $\pm$   0.02  &    1.8          & 0.56 $\pm$ 0.10      &  33.5           & 0.04 &    (2) \\
 C$_2$H$_5$CN                            &  27$_{13,14}$--26$_{13,13}$                                                                                              &  350       & 241.99710       &   0.12 $\pm$   0.02  &    3.2          & 0.40 $\pm$ 0.17      &  33.2           & 0.04 &    (2) \\
 C$_2$H$_5$CN                            &  27$_{7,21}$--26$_{7,20}$                                                                                                &  217       & 242.05249       &   0.23 $\pm$   0.02  &    4.0          & 0.97 $\pm$ 0.22      &  34.2           & 0.04 &    (2) \\
\enddata
\tablecomments{
(1) Blend of four C$_2$H$_5$CN lines with similar spectroscopic constants. 
(2) Blend of two C$_2$H$_5$CN lines with similar spectroscopic constants. 
\added{(3) Tentative detection. }
(4) Blend of four NH$_2$CHO lines with similar spectroscopic constants. 
(5) Blend of two NH$_2$CHO lines with similar spectroscopic constants. 
}
\end{deluxetable*}

\startlongtable
\begin{deluxetable*}{ l c c c c c c c c c }
\tablecaption{Line Parameters for HCOOH, H$_2$CCO, and c-C$_2$H$_4$O \label{tab_lines9}}
\tablewidth{0pt}
\tabletypesize{\scriptsize} 
\tablehead{
\colhead{Molecule}                       & \colhead{Transition}                                                             &       \colhead{$E_{u}$} &       \colhead{Frequency} &        \colhead{$T_{br}$} &     \colhead{$\Delta$$V$} &     \colhead{$\int T_{br} dV$} &       \colhead{$V_{LSR}$} &        \colhead{RMS} &       \colhead{Note} \\
\colhead{ }                              & \colhead{ }                                                                      &        \colhead{(K)} &           \colhead{(GHz)} &             \colhead{(K)} &          \colhead{(km/s)} &             \colhead{(K km/s)} &          \colhead{(km/s)} &        \colhead{(K)} &           \colhead{}
}
\startdata
 \textit{trans}-HCOOH                    &  12$_{1,12}$--11$_{1,11}$                                                                                                &   84       & 257.97501       &   0.52 $\pm$   0.02  &    5.5          & 3.00 $\pm$ 0.23      &  34.5           & 0.03 &    \nodata \\
 \textit{trans}-HCOOH                    &  15$_{4,12}$--14$_{4,11}$                                                                                                &  181       & 338.14384       &   0.42 $\pm$   0.03  &    5.0          & 2.26 $\pm$ 0.43      &  34.6           & 0.07 &    \nodata \\
 \textit{trans}-HCOOH                    &  15$_{3,13}$--14$_{3,12}$                                                                                                &  158       & 338.20186       &   0.36 $\pm$   0.03  &    4.4          & 1.70 $\pm$ 0.39      &  34.7           & 0.07 &    \nodata \\
 \textit{trans}-HCOOH                    &  15$_{4,11}$--14$_{4,10}$                                                                                                &  181       & 338.24882       &   0.35 $\pm$   0.04  &    2.8          & 1.04 $\pm$ 0.26      &  34.7           & 0.07 &    \nodata \\
 \textit{trans}-HCOOH                    &  15$_{3,12}$--14$_{3,11}$                                                                                                &  159       & 340.22910       &   0.31 $\pm$   0.04  &    5.8          & 1.92 $\pm$ 0.49      &  35.1           & 0.07 &    \nodata \\
 \textit{cis}-HCOOH                      &  13$_{0,13}$--12$_{1,12}$                                                                                                &   95       & 244.23510       &   0.33 $\pm$   0.02  &    1.6          & 0.57 $\pm$ 0.09      &  35.7           & 0.04 &    \nodata \\
\enddata
\tablecomments{
(1) Partial blend with SO$_2$. 
(2) Blend of two H$_2$CCO lines with similar spectroscopic constants. 
(3) Partial blend with C$_2$H$_5$OH. 
(4) Blend of two c-C$_2$H$_4$O lines with similar spectroscopic constants. 
}
\end{deluxetable*}

\clearpage

\section{Fitted spectra}
\begin{figure*}[tp] 
\begin{center} 
\includegraphics[width=17.0cm]{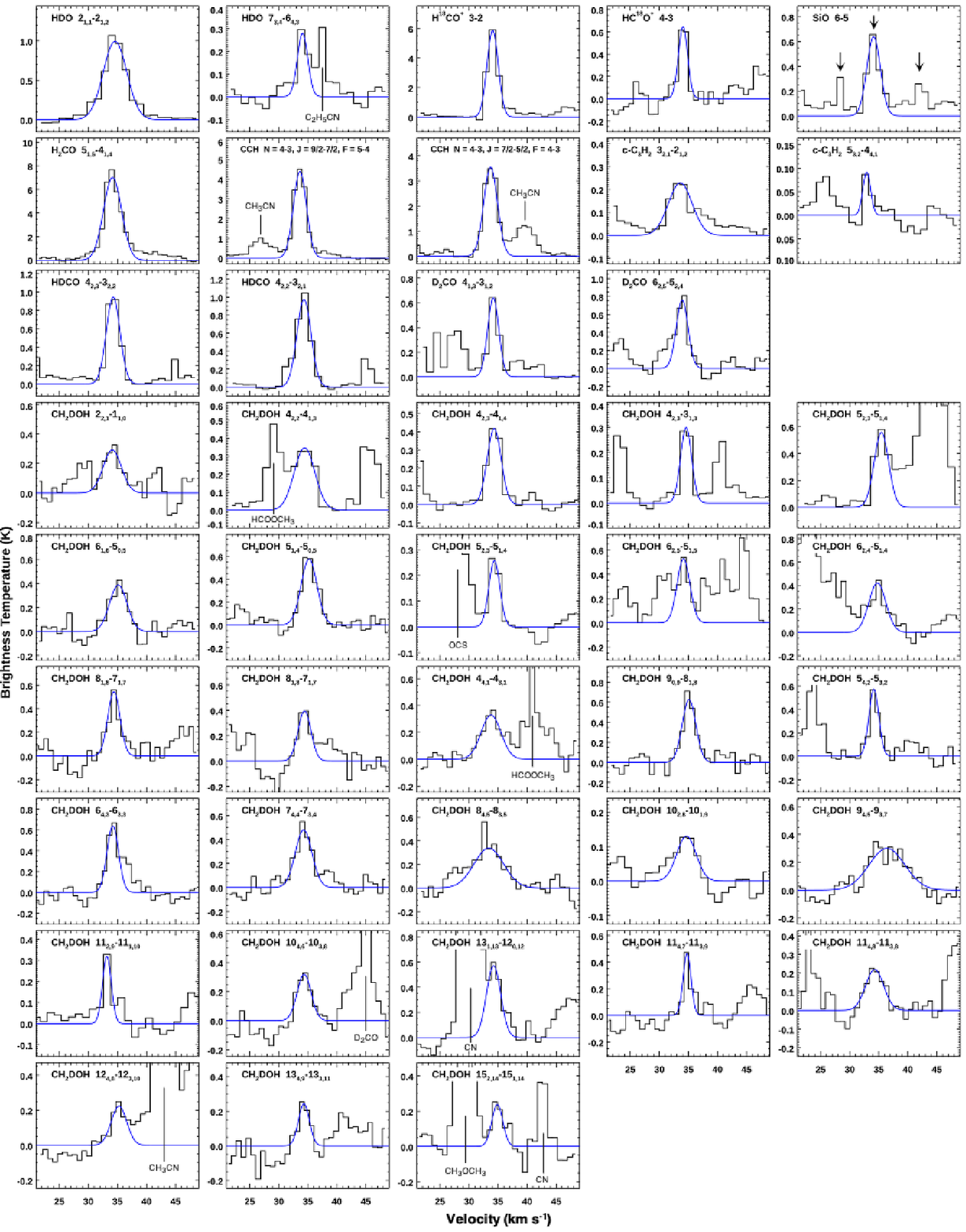} 
\caption{
ALMA spectra of the detected molecular emission lines. 
The blue lines represent fitted Gaussian profiles. 
For the molecules with multiple line detection, the spectra are sorted in ascending order of the upper state energy (the emission line with the lowest upper state energy is shown in the upper left panel and that with the highest energy is in the lower right panel). 
For SiO, the positions of primary and secondary peaks are indicated by arrows. 
}
\label{line_others}
\end{center}
\end{figure*}

\begin{figure*}[tp]
\begin{center}
\includegraphics[width=17.0cm]{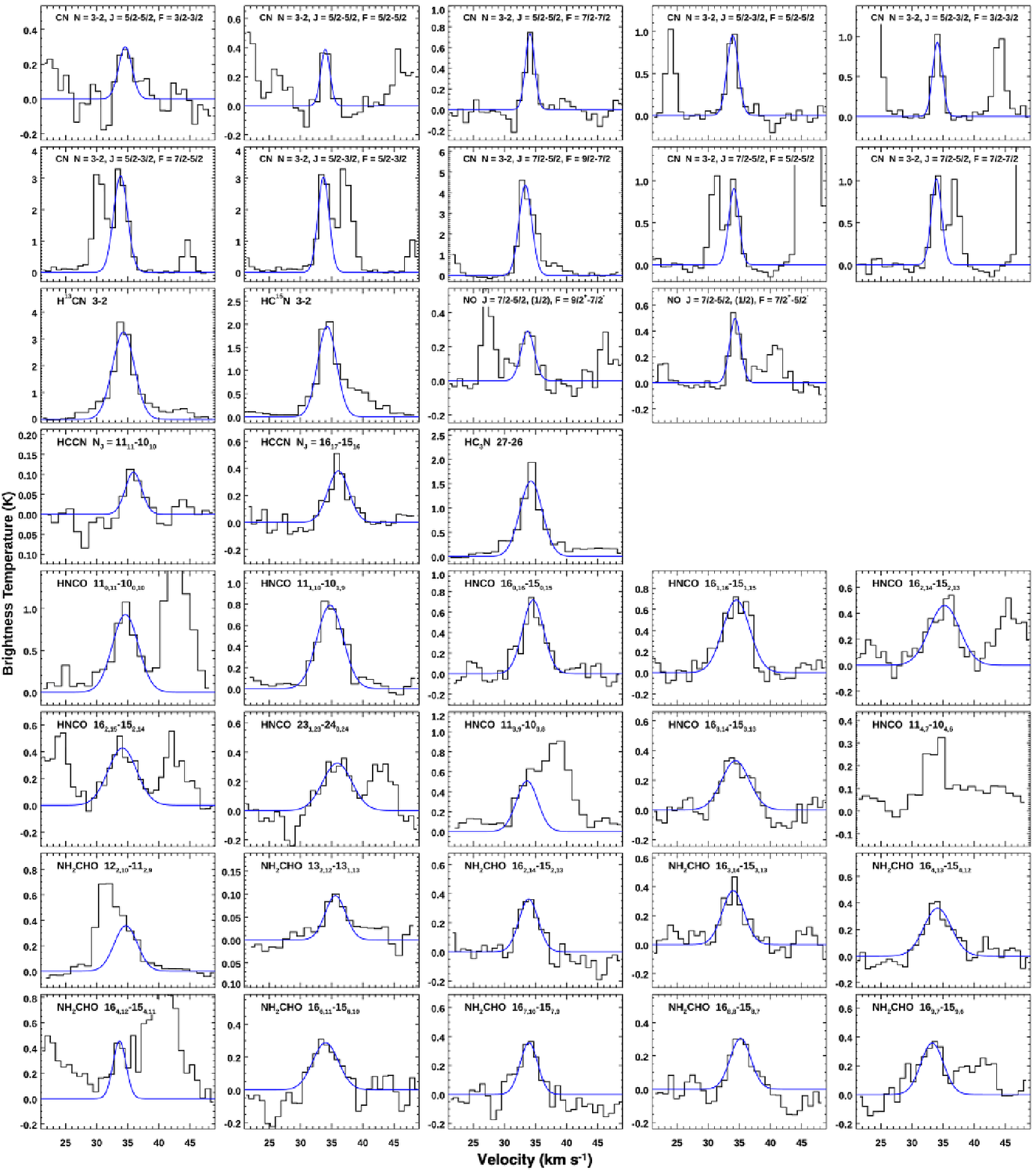}
\caption{
Same as in Figure \ref{line_others} but for nitrogen-bearing molecules. 
}
\label{line_Nbearing1}
\end{center}
\end{figure*}

\begin{figure*}[tp]
\begin{center}
\includegraphics[width=17.0cm]{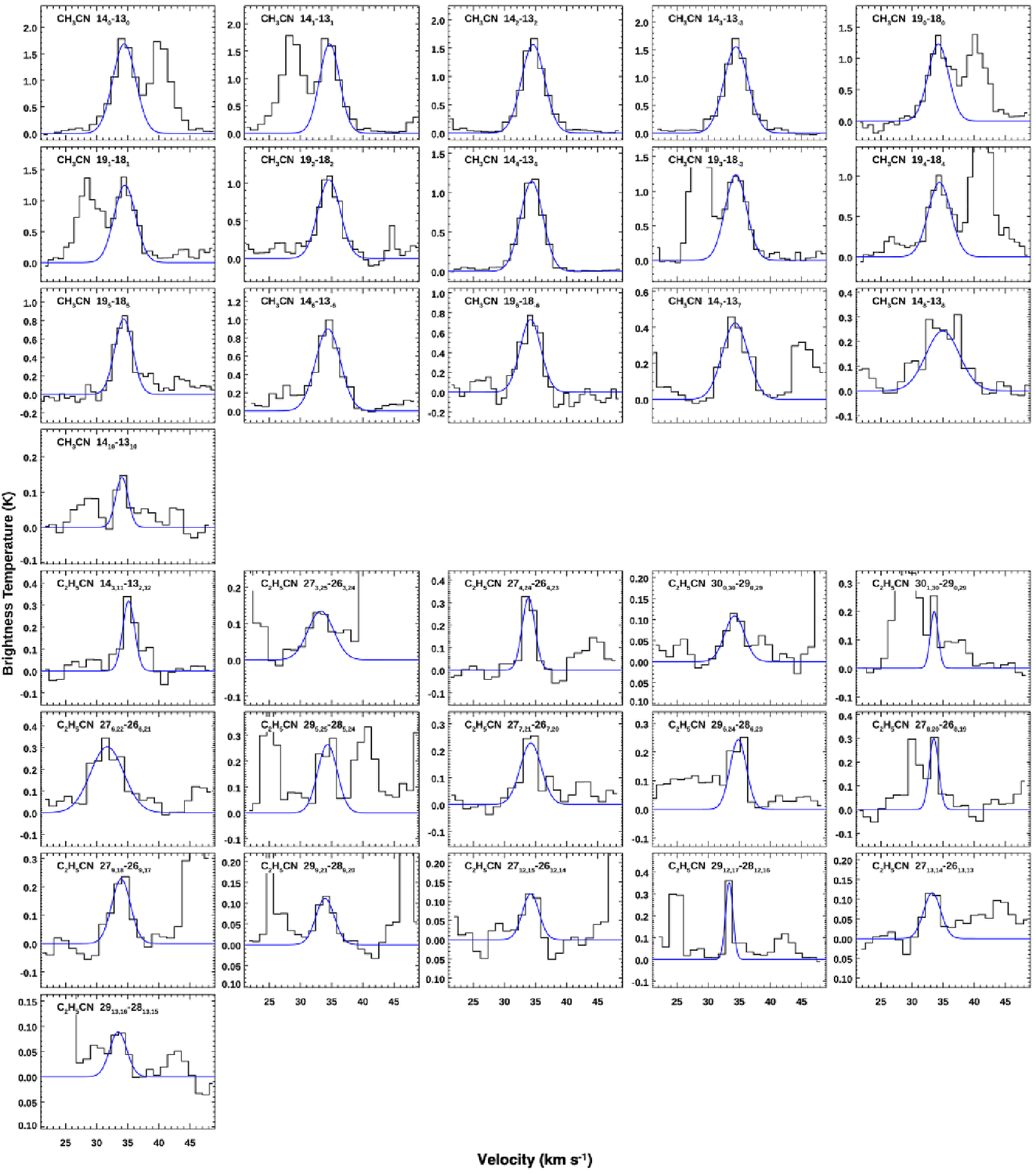}
\caption{
Same as in Figure \ref{line_others} but for nitrogen-bearing molecules (continued). 
}
\label{line_Nbearing2}
\end{center}
\end{figure*}

\begin{figure*}[tp]
\begin{center}
\includegraphics[width=17.0cm]{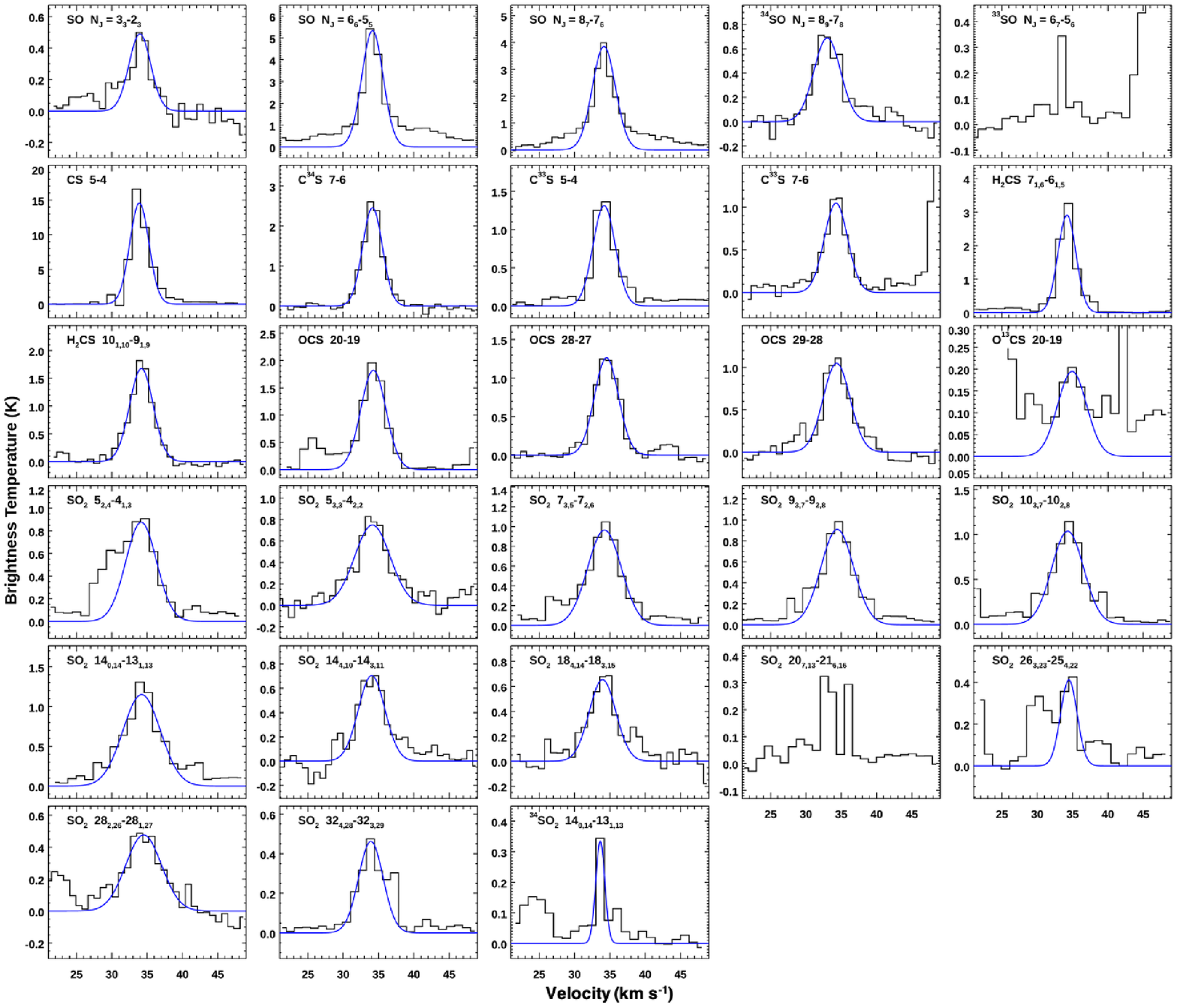}
\caption{
Same as in Figure \ref{line_others} but for sulfur-bearing molecules. 
}
\label{line_Sbearing1}
\end{center}
\end{figure*}

\begin{figure*}[tp]
\begin{center}
\includegraphics[width=17.0cm]{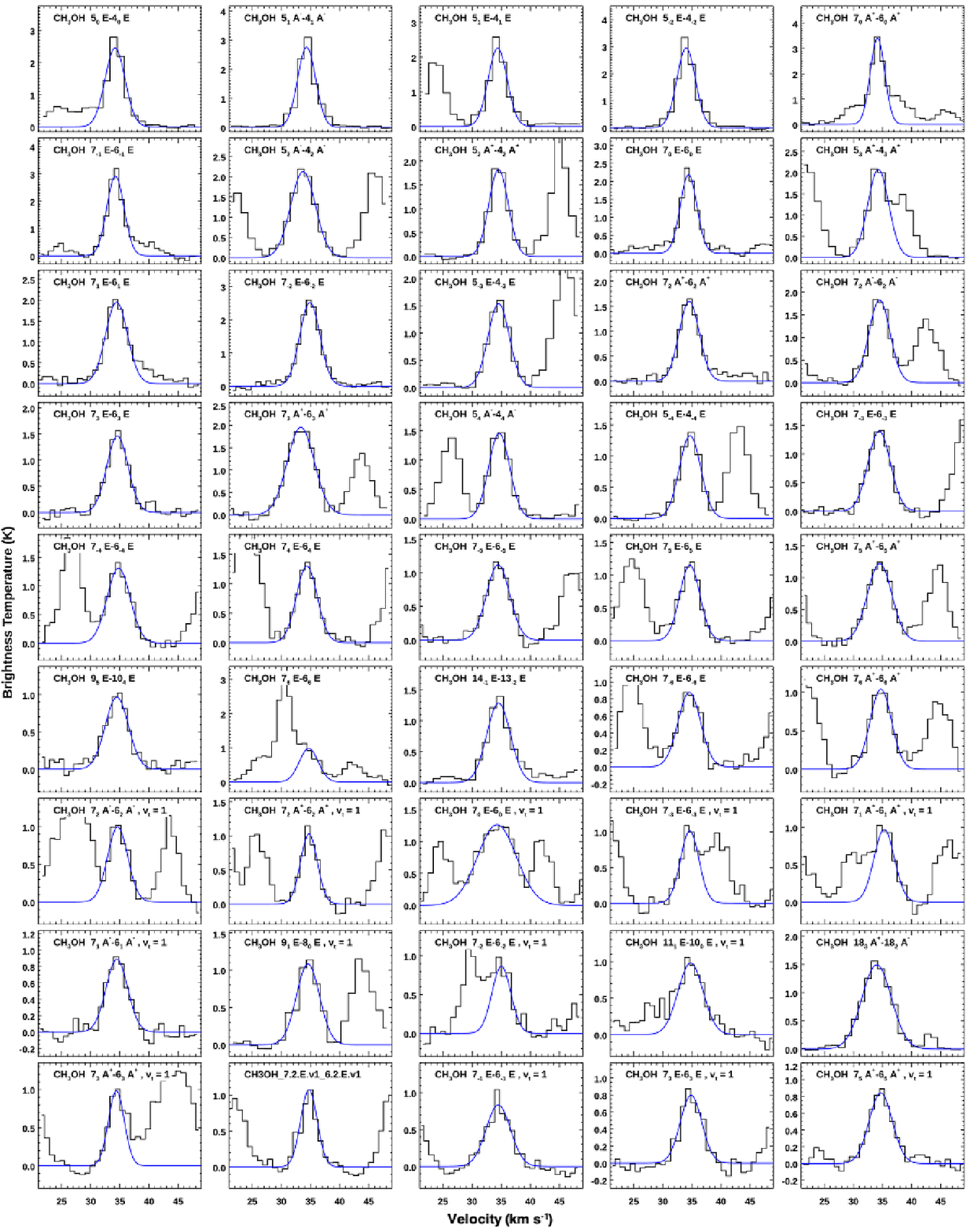}
\caption{
Same as in Figure \ref{line_others} but for CH$_3$OH. 
}
\label{line_CH3OH_1}
\end{center}
\end{figure*}

\begin{figure*}[tp]
\begin{center}
\includegraphics[width=17.0cm]{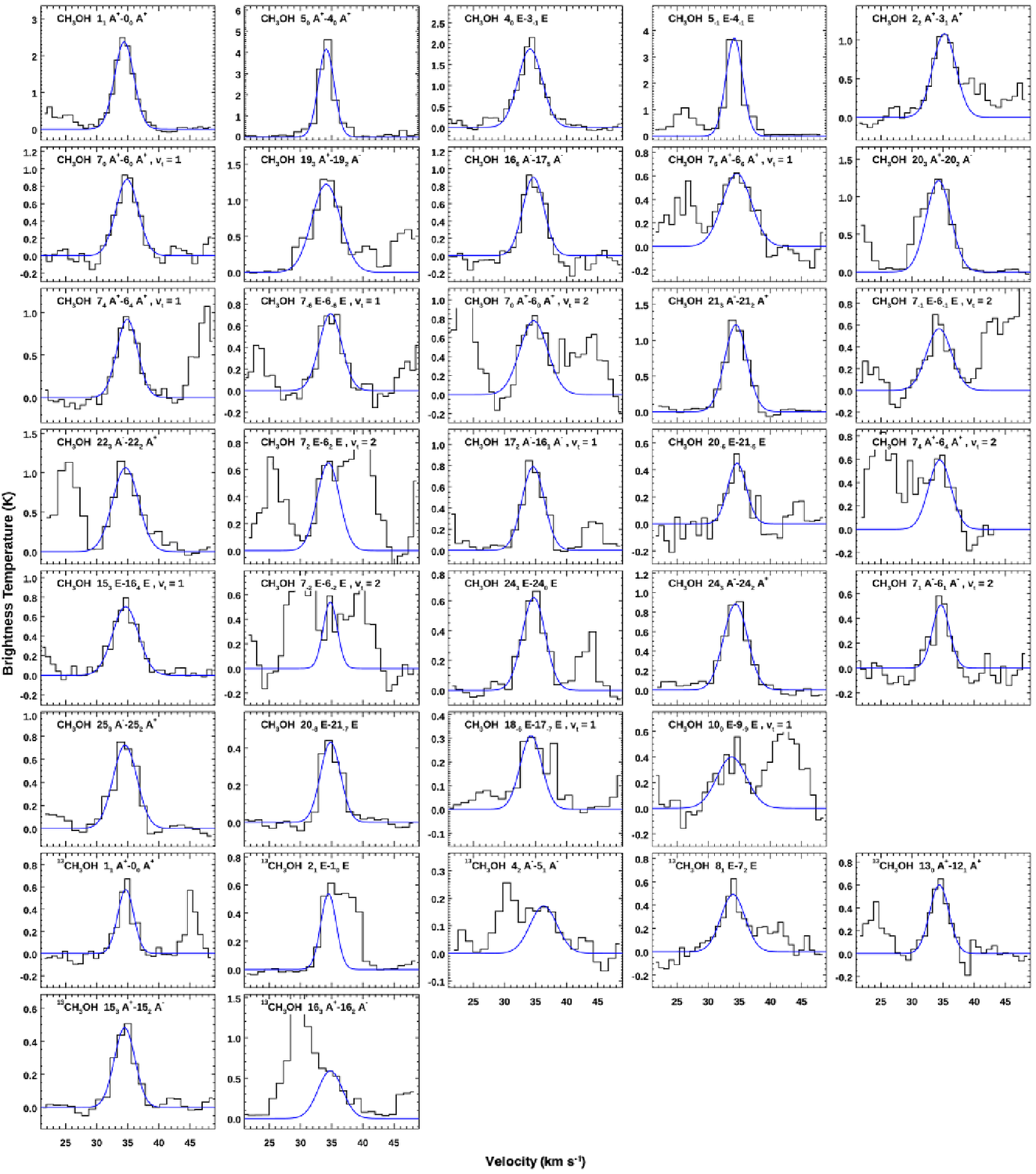}
\caption{
Same as in Figure \ref{line_others} but for CH$_3$OH (continued) and $^{13}$CH$_3$OH. 
}
\label{line_CH3OH_2}
\end{center}
\end{figure*}

\begin{figure*}[tp]
\begin{center}
\includegraphics[width=17.0cm]{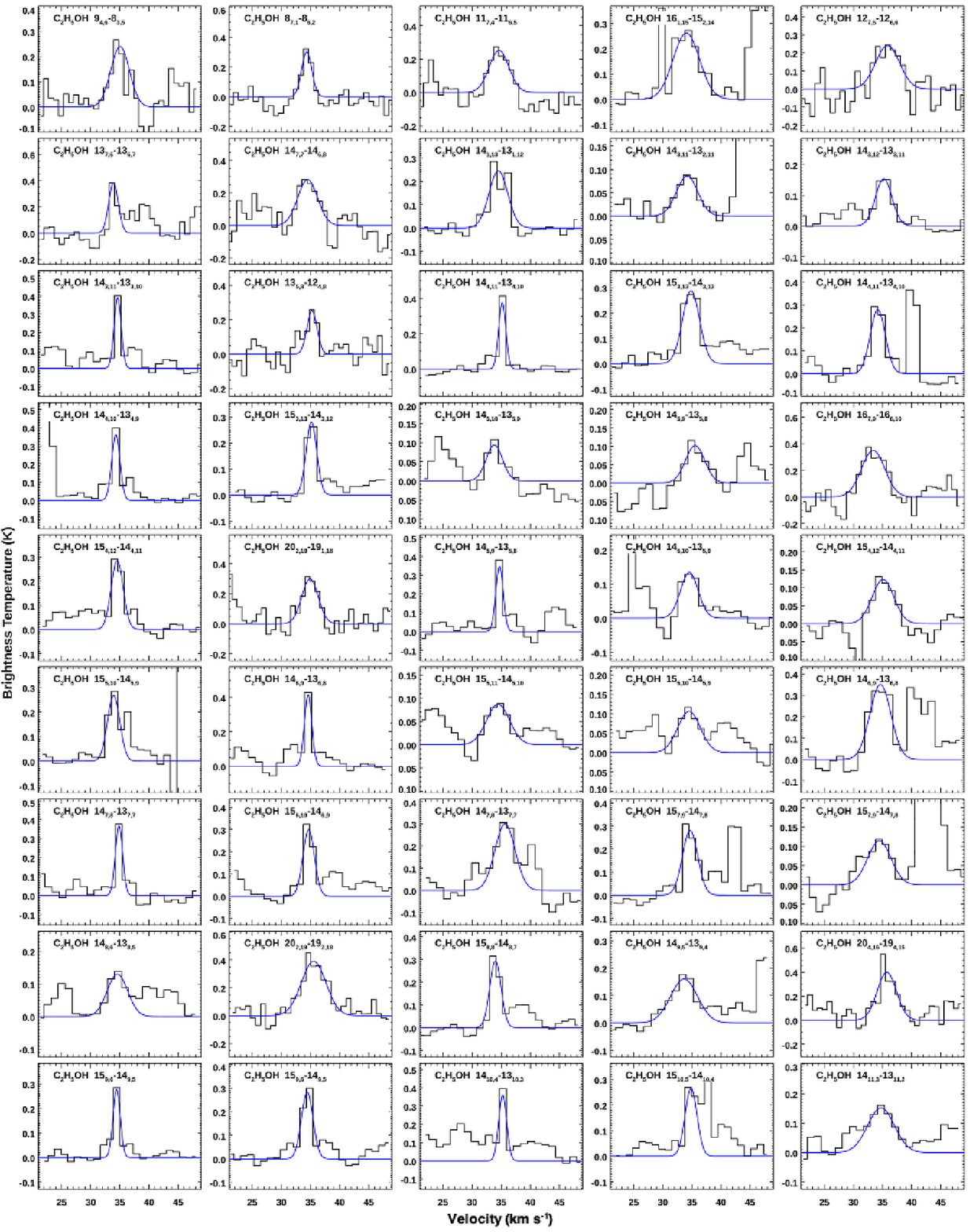}
\caption{
Same as in Figure \ref{line_others} but for C$_2$H$_5$OH. 
}
\label{line_C2H5OH}
\end{center}
\end{figure*}

\begin{figure*}[tp]
\begin{center}
\includegraphics[width=17.0cm]{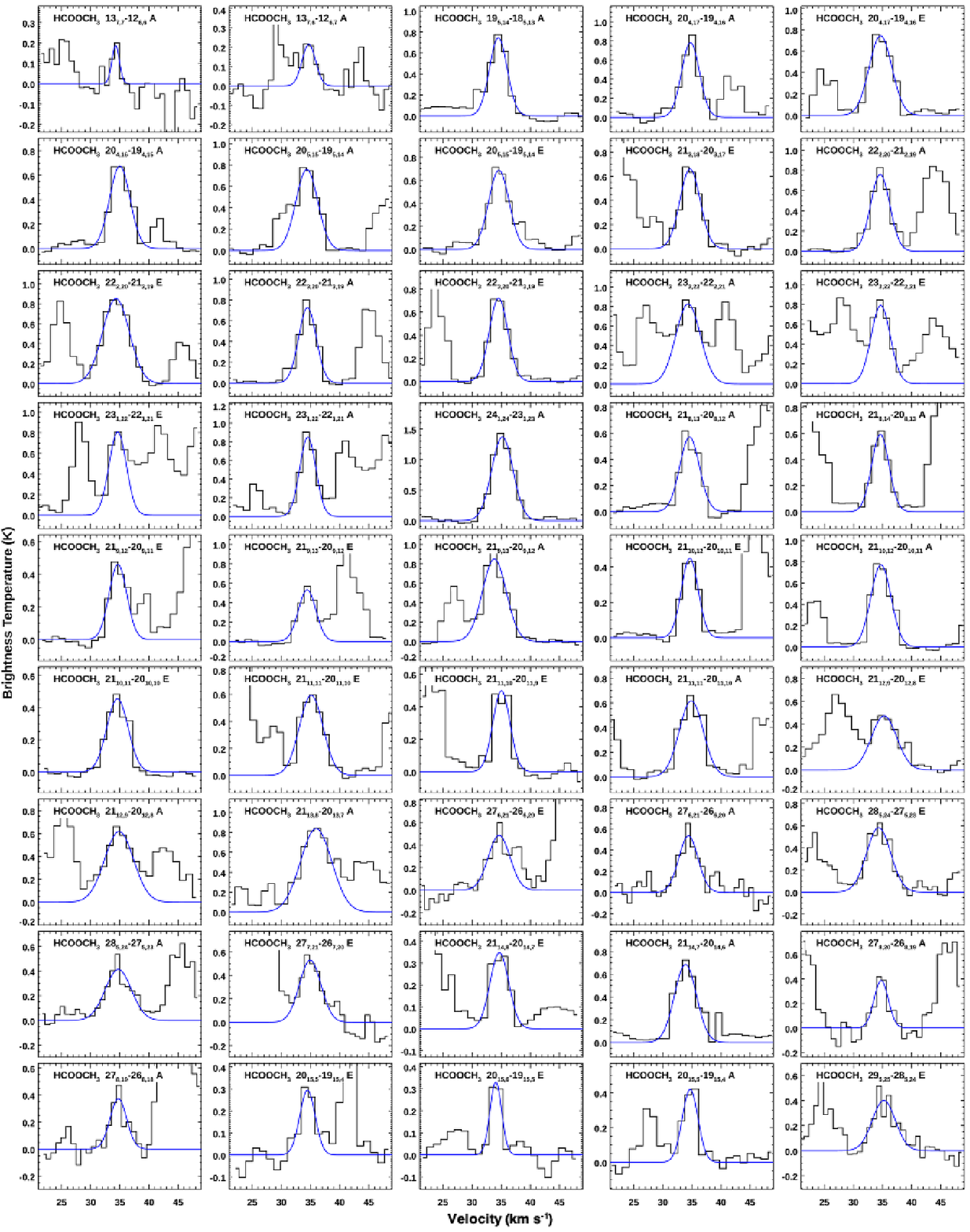}
\caption{
Same as in Figure \ref{line_others} but for HCOOCH$_3$. 
}
\label{line_HCOOCH3_1}
\end{center}
\end{figure*}

\begin{figure*}[tp]
\begin{center}
\includegraphics[width=17.0cm]{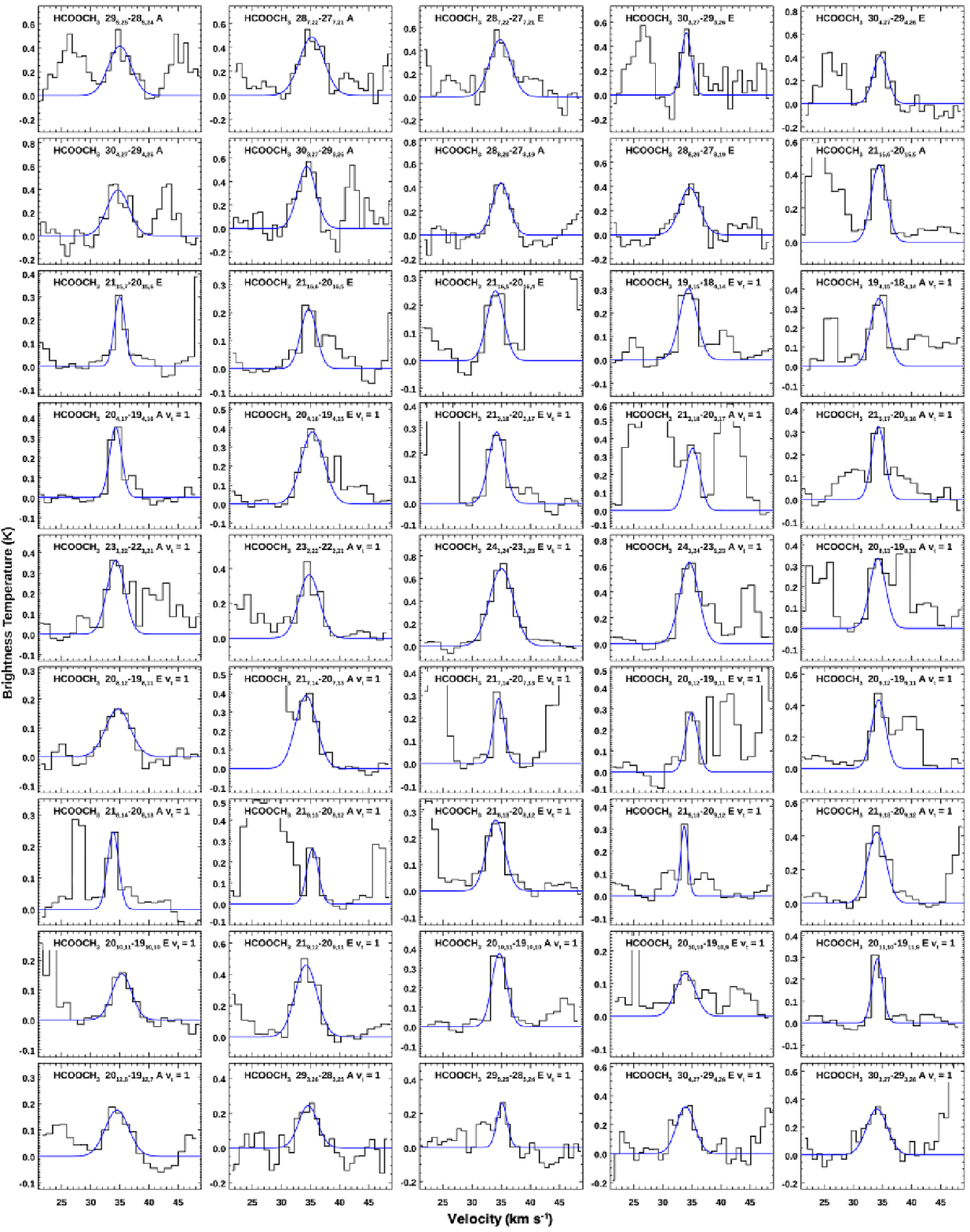}
\caption{
Same as in Figure \ref{line_others} but for HCOOCH$_3$ (continued). 
}
\label{line_HCOOCH3_2}
\end{center}
\end{figure*}

\begin{figure*}[tp]
\begin{center}
\includegraphics[width=17.0cm]{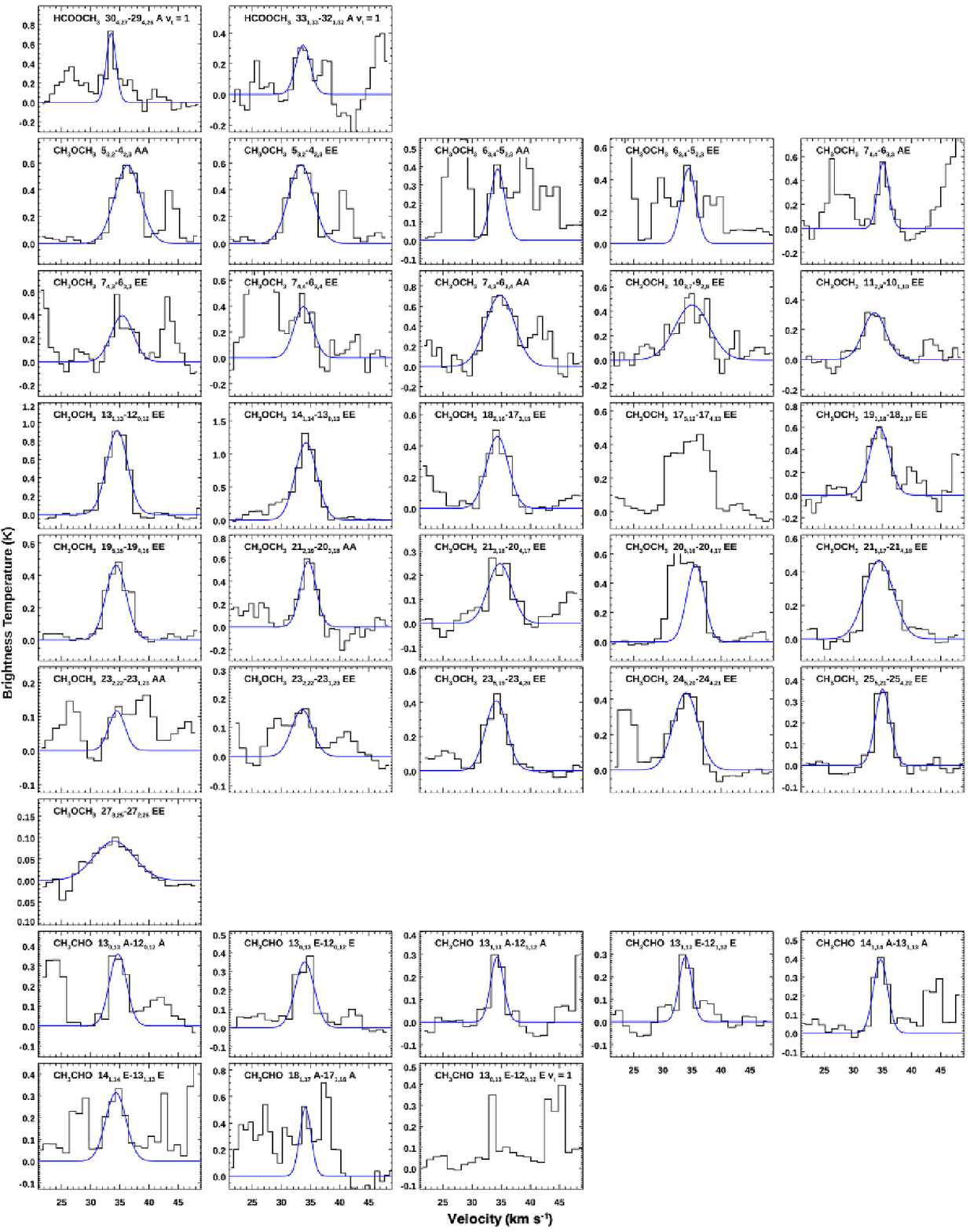}
\caption{
Same as in Figure \ref{line_others} but for HCOOCH$_3$ (continued), CH$_3$OCH$_3$, and CH$_3$CHO. 
}
\label{line_otherCOMs_1}
\end{center}
\end{figure*}

\begin{figure*}[tp]
\begin{center}
\includegraphics[width=17.0cm]{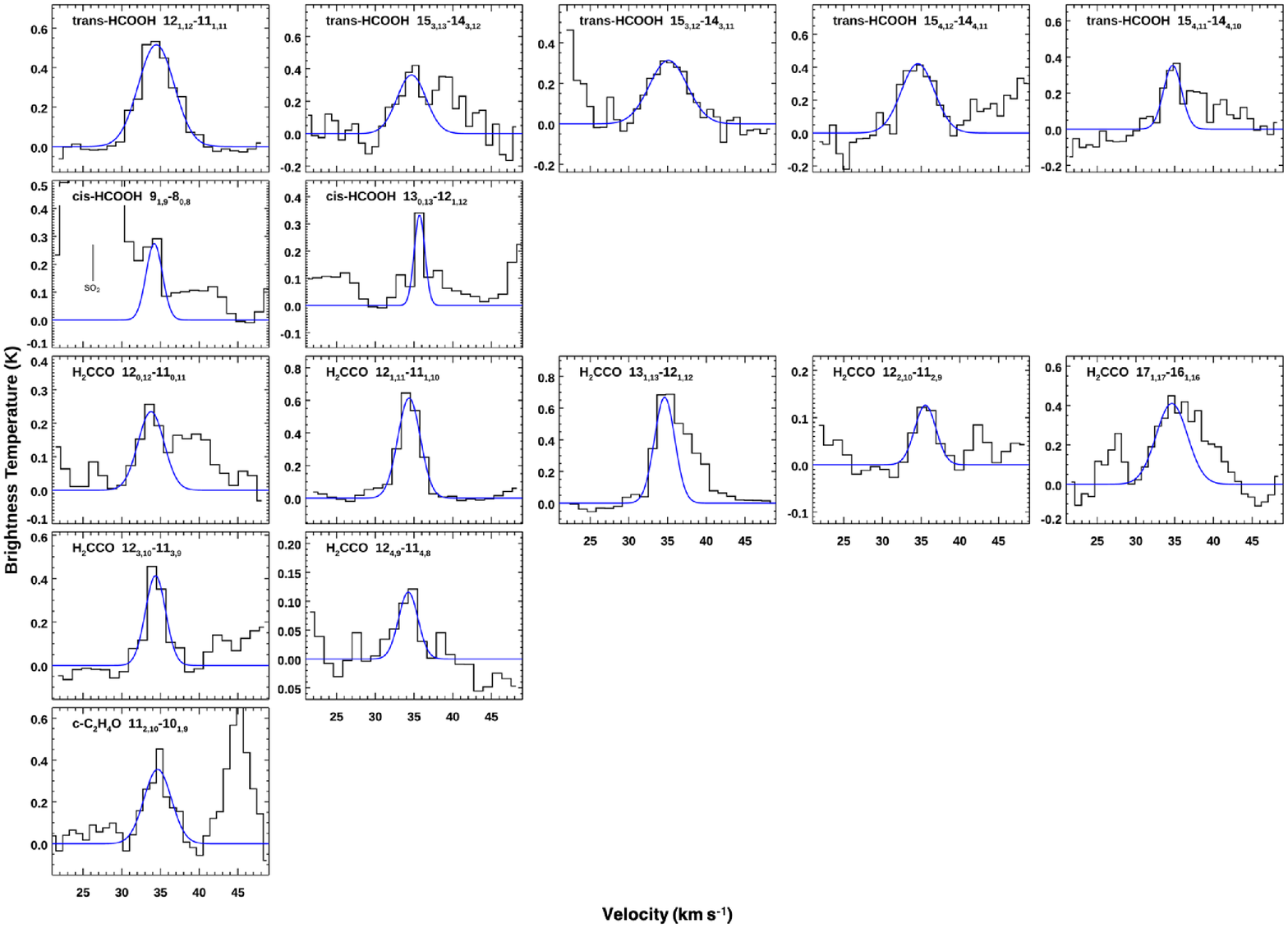}
\caption{
Same as in Figure \ref{line_others} but for HCOOH, H$_2$CCO, and c-C$_2$H$_4$O. 
}
\label{line_otherCOMs_2}
\end{center}
\end{figure*}

Figures \ref{line_others}--\ref{line_otherCOMs_2} show the results of the spectral line fitting (see Section \ref{sec_spc} for details). 

\end{document}